\documentclass[conference]{IEEEtran}
\IEEEoverridecommandlockouts

\usepackage{graphicx}
\usepackage{textcomp}
\usepackage{xcolor}
\usepackage[hyphens]{url}
\usepackage{hyperref}

\usepackage{float}
\usepackage{makecell}
\usepackage{algorithm} 
\usepackage[noend]{algpseudocode} 
\usepackage[capitalise]{cleveref}
\usepackage{tabularx}
\usepackage{listings}
\usepackage{enumitem}
\usepackage{multirow}
\usepackage{footnote}

\usepackage{pifont} %
\usepackage{makecell} %
\usepackage[frozencache]{minted}

\usepackage{subcaption} 

\def\BibTeX{{\rm B\kern-.05em{\sc i\kern-.025em b}\kern-.08em
    T\kern-.1667em\lower.7ex\hbox{E}\kern-.125emX}}
\begin{document}

\pdfpagewidth=8.5in
\pdfpageheight=11in

\title{Caerus: Prefetching by Thinking Globally and Acting Locally}

\author{ Jacky Wong\textsuperscript{1}\IEEEauthorrefmark{1}\IEEEauthorrefmark{2}\thanks{\IEEEauthorrefmark{1}Equal contribution.}\thanks{\IEEEauthorrefmark{2} Corresponding Author: jacky.wong20@imperial.ac.uk}, Jacob Alexandrou\textsuperscript{1}\IEEEauthorrefmark{1}, Yinting Huang\textsuperscript{2}, Luke Panayi\textsuperscript{1} \\ Lev Mukhanov\textsuperscript{3}, Sam Ainsworth\textsuperscript{3}, Paul H. J. Kelly\textsuperscript{1} \\ \\

\textsuperscript{1}Imperial College London, United Kingdom \quad \textsuperscript{2}University of Cambridge, United Kingdom\\ \textsuperscript{3}IMEC, United Kingdom  

\thanks{This work has been partially funded by UKRI grant EP/Y020499/1. }
}

\lstset{
  language=C,
  columns=flexible,
  basicstyle={\small\ttfamily},
  numbers=none,
  numberstyle=\tiny\color{gray},
  keywordstyle=\color{blue},
  commentstyle=\color{dkgreen},
  stringstyle=\color{mauve},
  breaklines=true,
  breakatwhitespace=true,
  tabsize=2
}

\definecolor{pcYellow}{HTML}{FFD54F}  %
\definecolor{pcGreen}{HTML}{81C784}   %
\definecolor{pcOrange}{HTML}{FFB74D}  %
\definecolor{pcBlue}{HTML}{7986CB}    %
\definecolor{pcRed}{HTML}{E57373}     %
\definecolor{pcGray}{HTML}{D0D0D0}    %

\definecolor{fig6blue}{HTML}{007FFF} 
\definecolor{fig6green}{HTML}{009900}
\definecolor{fig6red}{HTML}{CC0000}

\newcommand{\sam}[1]{\textbf{\textcolor{blue}{\textit{Sam:} #1}}}

\newcommand{\jacky}[1]{\textbf{\textcolor{orange}{\textit{Jacky:} #1}}}
\newcommand{\paul}[1]{\textbf{\textcolor{purple}{\textit{Paul:} #1}}}
\newcommand{\jacob}[1]{\textbf{\textcolor{green}{\textit{Jacob:} #1}}}
\newcommand{\lev}[1]{\textbf{\textcolor{pink}{\textit{Lev:} #1}}}
\newcommand{\justin}[1]{\textbf{\textcolor{green}{\textit{Justin:} #1}}}

\newcommand{\luke}[1]{\textbf{\textcolor{yellow}{\textit{Luke:} #1}}}

\newcommand{\todo}[1]{\textbf{\textcolor{red}{\textit{Todo:} #1}}}

\newcommand{\revision}[1]{#1}
\newcommand{\toalter}[1]{\textcolor{purple}{#1}}

\newcommand{\bench}[1]{\texttt{#1}}
\newcommand{\ignore}[1]{}

\newcommand{\gspace}[1]{}

\maketitle

\begin{abstract}

Modern processors rely heavily on data prefetching to bridge the growing performance gap between computation and memory accesses. In offset prefetching, prefetches are issued at an address delta relative to the most recently observed access, selecting deltas that have exhibited high coverage in the recent access-stream history. The Best-Offset Prefetcher (BOP) exemplifies this approach by selecting a single uniformly-issued global offset that delivers timely prefetches. However, its lack of context-sensitive accuracy tracking results in significantly increased DRAM traffic and sometimes disappointing performance. Conversely, Berti, the state-of-the-art offset prefetcher, achieves impressive prefetching performance and accuracy. However, its Program-Counter (PC)-local approach to extracting offsets limits its ability to capture global memory patterns, meaning there are many scenarios where simpler global offset prefetchers, such as BOP, outperform it. 

Observing the limitations of contemporary offset prefetchers, we present Caerus: an offset prefetcher that employs PC-local filtering on multiple non-overlapping globally derived offsets. Caerus extracts multiple offsets from the global memory trace and maximizes their combined coverage by training each offset only on accesses not already covered by previously learned offsets. The accuracy of each extracted offset is tracked per-PC based on whether it can cover future accesses for any PC. This allows Caerus to address complex delta patterns while maintaining high accuracy. In SPECspeed 2017, Caerus gains a geomean speedup of 8.3\%, outperforming Berti by 2.0\% while achieving low hardware and DRAM-traffic overhead.

\end{abstract}


\section{Introduction}

Modern processors are increasingly constrained by memory latency. As CPU cores become faster and more parallel, the gap between computation speed and memory access time is expanding. This is particularly problematic in memory-intensive workloads where the cost of waiting for memory transactions dominates execution time. 

To bridge this gap, almost all modern CPUs employ data prefetchers that speculate on future memory accesses and fetch addresses before they are explicitly requested. Offset prefetchers~\cite{Sandbox,berto,he_chimera_2024,Hyperion,shakerinava_multi-lookahead_2019} have emerged as the preferred method of spatial prefetching, with the state-of-the-art being Berti~\cite{navarro-torres_berti_2022}. Their practical relevance is further demonstrated by the adoption of the Best-Offset Prefetcher (BOP)~\cite{michaud_best-offset_2016} in real systems~\cite{bruce_arm_2023,Arm-N2-Hotchips}. 

Offset prefetchers trigger timely prefetches by looking far enough back in the address stream, and can cover complex repeating patterns such as (+2, +3, +2, +3) by learning a compound offset such as +5, rather than being limited to a single fixed stride (see \cref{intro:offset_prefetching}). These factors significantly enhance the coverage and timeliness of offset prefetchers. Furthermore, since offset prefetchers do not obey precise ordering constraints, they are invariant to the negative effects of out-of-order execution.

\begin{figure}[t]
    \centering
\resizebox{\columnwidth}{!}{%
    \includegraphics{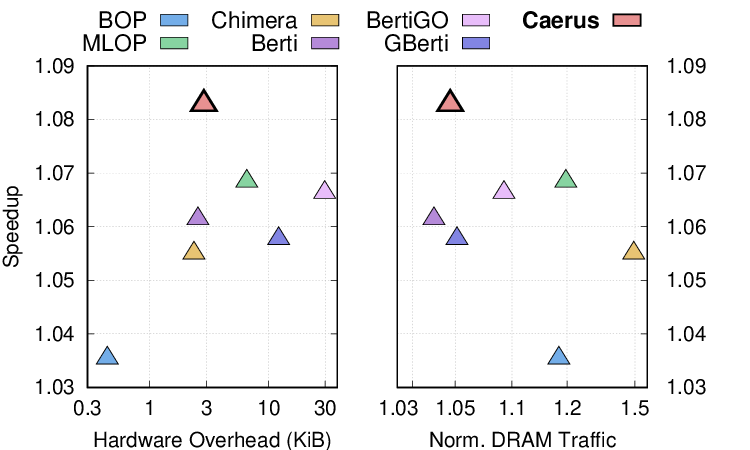}
}    \caption{
 Geomean speedup (SPEC2017) vs. hardware/DRAM-traffic overhead (log-scale) of BOP~\cite{michaud_best-offset_2016}, MLOP~\cite{shakerinava_multi-lookahead_2019}, Chimera~\cite{he_chimera_2024}, Berti~\cite{navarro-torres_berti_2022}, BertiGO~\cite{Singh2026BertiGO}, GBerti~\cite{Posluns2026GlobalBerti} and Caerus. Caerus achieves the highest performance while incurring low DRAM traffic and hardware cost. }
    \label{fig:speedup-combined}
\end{figure}

However, we observe that Berti~\cite{navarro-torres_berti_2022,berto}, the current state-of-the-art design, has coverage limitations that harm its potential. This is due to its completely PC-localized training and filtering: an offset trained at one PC can never cover accesses from another PC (see \cref{ssec:berti_limitations}). Recent top DPC4~\cite{dpc4_2026} entrants BertiGO~\cite{Singh2026BertiGO} and GBerti~\cite{Posluns2026GlobalBerti} do not fix these underlying issues, with the winner, BertiGO, only improving on Berti by 0.5\% (\cref{fig:speedup-combined}). Composite prefetchers such as Chimera~\cite{he_chimera_2024} and Hyperion~\cite{Hyperion} acknowledge this inherent drawback and incorporate heuristics that employ orthogonal prefetchers. Hyperion derives offsets localized to cover entire individual regions/pages~\cite{he_chimera_2024,Hyperion}, while Chimera incorporates a BOP-like~\cite{michaud_best-offset_2016} prefetcher that uses one offset to cover all accesses~\cite{he_chimera_2024}. However, both of these designs fail to realize and exploit an important property: that the performant offsets at a given PC may effectively cover the access of an entirely different PC. 

BOP~\cite{michaud_best-offset_2016} avoids this limitation by considering offsets derived from a global scope. However, issuing the same offset prefetch at every PC ignores the local execution context and is often inaccurate~\cite{navarro-torres_berti_2022}. This limits BOP to issuing just one offset, in order to not increase inaccuracy further. MLOP~\cite{shakerinava_multi-lookahead_2019} issues from many offsets by trading off timeliness and coverage with each, but suffers from very high DRAM traffic, in turn harming performance (see \cref{ssec:bop_limitations}). 

\ignore{At the same time, the end of Dennard scaling means energy efficiency and power have become vital metrics as mobile devices, to desktop processors, to state-of-the-art data centers hit a "Power Wall"~\cite{power_wall}. Hardware vendors have thus become less tolerant~\cite{AMD‑30x25‑2024,qualcomm_power_efficient_technologies_2021} of inaccurate prefetchers that increase DRAM traffic, and as a result energy consumption/temperature~\cite{EnergyPF,4771791,10.1145/3224419,Feedback}, without getting commensurate performance gain. Moreover, in mobile devices, redundant SoC and DRAM traffic can also degrade the performance of other components, such as GPUs and NPUs~\cite{arm_mobile_memory_whitepaper_2016,9211564}. Thus, memory traffic has become a central target metric in designing the prefetchers beyond performance. }

Given the limitations of contemporary offset prefetchers, we develop Caerus: a novel hybrid methodology that incorporates the granular per-PC accuracy tracking strategy of Berti, enabling a variable number of distinct offsets to be issued at different PCs, while preserving the comprehensive, global training scope of BOP. This requires several new insights:

\begin{itemize}[leftmargin=*]
\item  We develop a \textit{next-best offset} selection algorithm to choose the set of offsets that collectively maximizes coverage by minimizing wasteful redundancy between distinct offsets that cover the same loads. This handles the problem of \textit{overlap}:
the best offset(s) at PC A might cover all the same loads
as the best offset(s) at PC B, leaving accesses at PC C not
prefetched at all.
\item We then apply a filter on these overlap-resistant offsets based on their relevance in the load's PC-local context, allowing Caerus to achieve both high coverage and accuracy. 
\item Finally we introduce a practical method for tracking timeliness inspired by BOP but generalized to complex conditional, multi-offset prefetches, avoiding Berti's complex cache timeliness tags.
\end{itemize}

\noindent Caerus achieves an  8.3\% speedup in SPECspeed 2017, outperforming all of its competitors by a wide margin (\cref{fig:speedup-combined}) while being highly frugal. Caerus achieves comparable DRAM traffic and hardware size to the original Berti~\cite{navarro-torres_berti_2022}, while outperforming it by 2.0\%. It also outperforms the DPC4 winner, BertiGO~\cite{Singh2026BertiGO}, by 1.5\% while consuming only a tenth of the hardware resources and half the excess DRAM traffic. 

\section{Motivation}
\label{ssec:motivation}

Existing offset prefetching algorithms differ in how they gather and filter offsets, leading to various trade-offs across performance metrics. This section analyzes the performance characteristics and gaps of three prevailing offset prefetchers.

\begin{figure}
    \centering
    \includegraphics[width=1\linewidth]{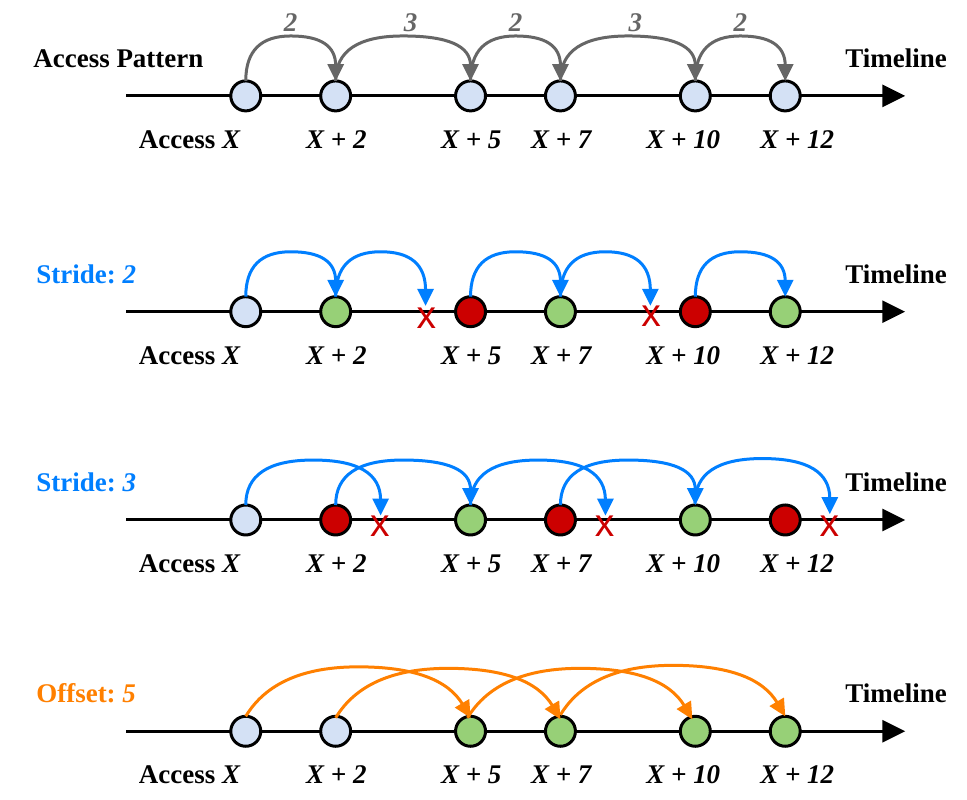}
    \caption{Illustration of offset prefetching. Two competing adjacent stride patterns +2 and +3 are present, but neither is consistently correct. Looking back further in the access history reveals that an offset access at +5, a combination of +2 and +3, will always give correct prefetches.}
    \label{intro:offset_prefetching}
\end{figure}

\paragraph{Offset Prefetching} Offset prefetching~\cite{Sandbox,michaud_best-offset_2016} can be seen as a generalization of stride prefetching~\cite{stride} (\cref{intro:offset_prefetching}). Stride prefetching analyzes adjacent accesses and therefore works well when address deltas are constant, but struggles when access patterns alternate. Offset prefetching instead looks further back in the access stream to identify more effective compound deltas. In \cref{intro:offset_prefetching}, analyzing adjacent accesses alone will yield a stride of 2 or of 3, each having 50\% accuracy and 50\% coverage. An offset of 5 will instead always cover a correct future access, with 100\% coverage and accuracy.

\paragraph{Best-Offset Prefetching} The Best Offset Prefetcher (BOP) \cite{michaud_best-offset_2016} is the winner of the 2nd Data Prefetching Championship (DPC2). It scores a predetermined offset list against the processor's global memory trace -- with ``global'' meaning the Program Counter (PC) of the each load instructions is not considered -- and extracts one ``best offset''. On cache miss or prefetch hit, BOP issues a prefetch to the miss address + the best offset. The scoring mechanism updates continuously, replacing the current offset if a better candidate is found or turning off prefetching if no good candidates exist. 

\paragraph{Multi-Lookahead Offset Prefetching} The Multi-Lookahead Offset Prefetcher (MLOP) \cite{shakerinava_multi-lookahead_2019} improves upon BOP by introducing multiple global offsets. For every offset, MLOP considers multiple levels of score where each level corresponds to a specific lookahead. MLOP uses these different lookaheads to balance the trade-off between prefetch timeliness (whether an issued prefetch arrives before the target demand request) and prefetch coverage. In practice, MLOP is often very performant but suffers from high induced DRAM traffic and low prefetch accuracy. 

\paragraph{Berti} Berti \cite{navarro-torres_berti_2022} is a state-of-the-art, accurate offset prefetcher. Its foundation lies in the belief that offsets should always be timely and PC-local to precisely capture the local execution context. Berti defines local offsets as the difference in cache line addresses between any two demand accesses issued from the same instruction (identified by their PC). Timely prefetches, or prefetches that successfully arrive in the cache before their predicted demand access, are identified by recording the latency of cache refills. Hence, Berti trains on a filtered view of the global memory trace restricted to timely and PC-local accesses, allowing it to aggressively and dynamically derive offsets. The larger BertiGO~\cite{Singh2026BertiGO} adds per-region filtering and tracking of multiple recent PCs. %

\ignore{Berto \cite{berto} revises Berti's structure sizes, introduces a new low-cost method for approximating timeliness, and incorporates improved hash functions. In this work, we use the Berti (High Performance) variant provided by Berti’s authors. Berti combines the most effective features of Berti and Berto (explicit timeliness tracking, enhanced hash functions, and updated hardware sizing) to achieve higher performance than either design.}

\begin{figure}[t]
    \centering
    \includegraphics[width=0.85\columnwidth]{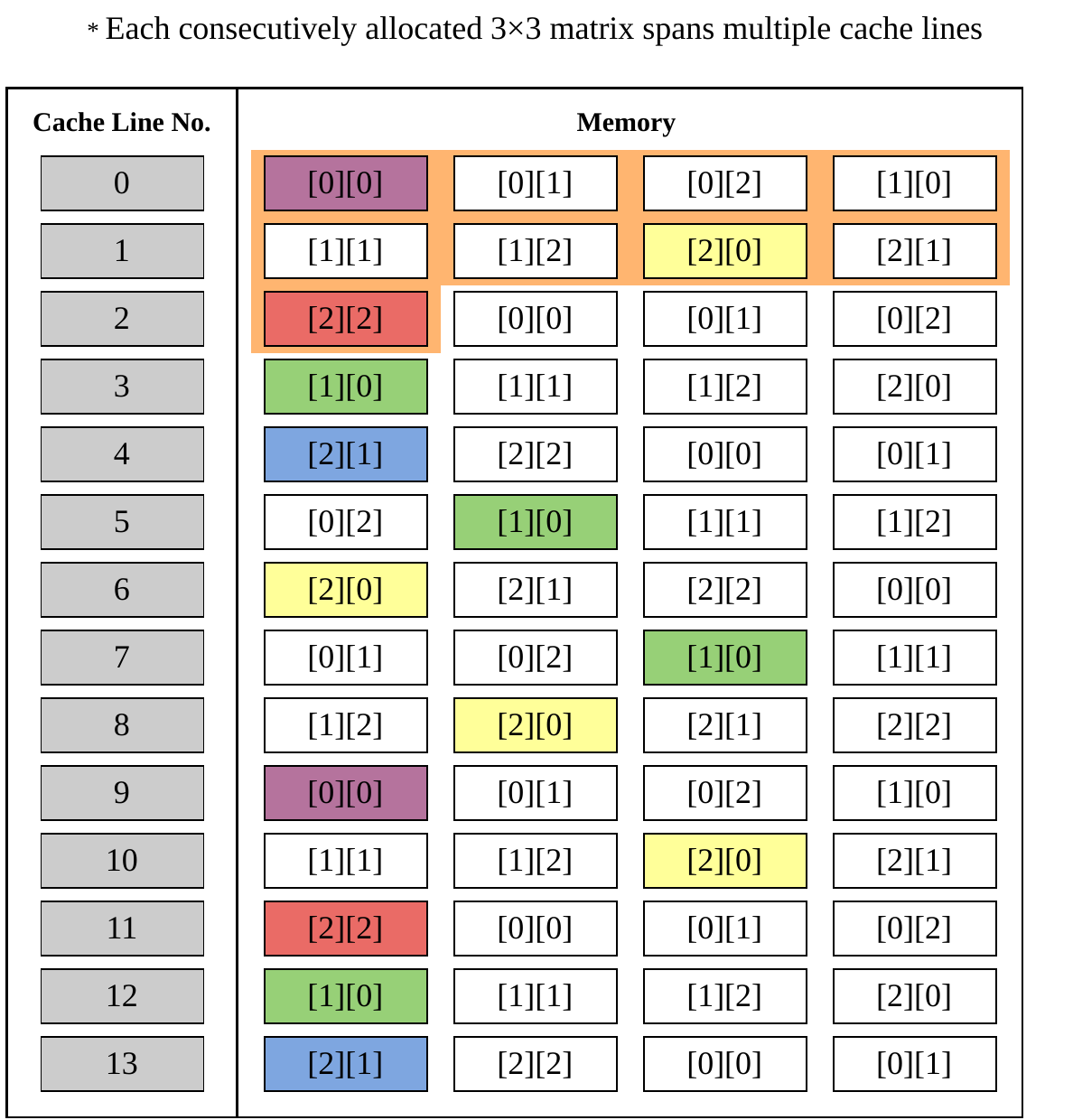} 
    \begin{lstlisting}
void mult_adj_su3_mat_vec(su3_matrix *a, su3_vector *b, su3_vector *c ){
  register double t,ar,ai,br,bi,cr,ci;
  for(int i=0;i<3;i++){
    for(int j=0;j<3;j++){
      ar=a->e[j][i].real; ai=a->e[j][i].imag;
      br=b->c[j].real; bi=b->c[j].imag;
      cr=ar*br; t=ai*bi; cr += t;
      ci=ar*bi; t=ai*br; ci -= t;
    } } }
\end{lstlisting}
    \caption{{Illustration of a cache-line access pattern from the \bench{milc} benchmark (SPEC CPU2006). After loop unrolling, each of the nine matrix elements is loaded at a distinct PC. The figure shows how elements of consecutively allocated 3×3 complex matrices map onto 64-byte cache lines. Because each matrix occupies 144 bytes, successive matrices begin at shifting cache-line offsets, and the PC that first touches each line (highlighted, distinct colors denote distinct PCs) varies.}
   }
    \label{fig:milc}
\end{figure}

\subsection{Limitations to PC-Local Fine-Grain Training}
\label{ssec:berti_limitations}

\begin{figure}
        \centering
    \subfloat[Speedup]{
    \includegraphics[width=\columnwidth]{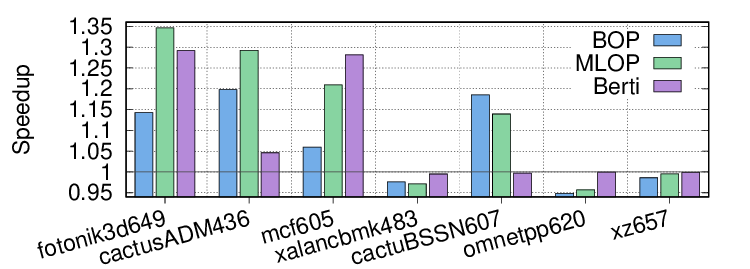}
    \label{motivation:bertibopipc}
   }
   
       \subfloat[Normalized DRAM Traffic]{
  \includegraphics[width=\columnwidth]{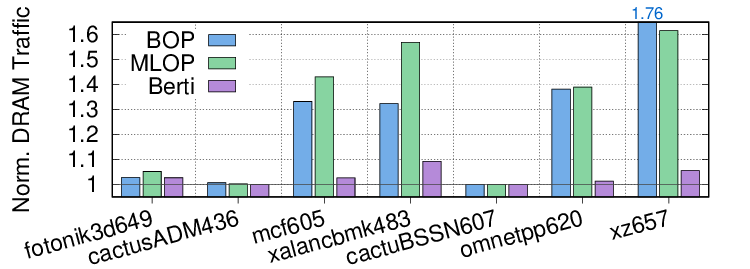}
    \label{motivation:bertibopdram}
   }
       \caption{Speedup and DRAM traffic of BOP, MLOP, and Berti in a selection of SPEC benchmarks.}
       \label{motivation:bertibop}
\end{figure}

Despite Berti's impressive performance, comparing the performance of BOP, MLOP and Berti on individual workloads reveals that Berti often fails to find performant offsets (see \bench{cactusADM436} and \bench{cactuBSSN607} in \cref{motivation:bertibopipc}). This is caused by Berti's focus on offsets strictly derived from a PC-local context, which stops it from recognizing prefetching opportunities that only emerge from analyzing global execution.

{\Cref{fig:milc} illustrates a seemingly innocuous example that exposes the limitations of PC-local offset training. \texttt{Mult\_adj\_su3\_mat\_vec}, a critical kernel in the \bench{milc} benchmark from SPEC CPU2006, multiplies a 3×3 complex-valued matrix by a vector. The inner loop iterates over the row index, and for each column accesses rows 0, 1, and 2, traversing the row-major matrix in column-major order. }

{After unrolling the loop, the compiler pairs each element's real and imaginary parts into a single 128-bit load, producing a straight-line sequence of 9 loads at 9 distinct PCs. When the call site iterates through an array of matrices, the aggregate access pattern is a simple sequential stream that touches every cache line in order. From the perspective of any individual load PC, however, the pattern is far less regular. Each matrix occupies $9\times16=144$ bytes, which is not a multiple of a 64-byte cache line. Successive matrices therefore begin at shifting offsets within their cache lines, and each cache line is touched by four of the nine load PCs. Crucially, the PC that touches a given cache line first in program order, and therefore incurs the initial cache miss, varies from one matrix to the next.}

Of the 9 PC-isolated loads, each cache line is accessed by four. A combination of the column-major access order, and cache-line misalignment, means that the first PC to access the line in program order, and the one that therefore observes the miss, is highly varied. Again assuming program order, the PCs that access (0,0), (2,1) and (2,2) always see a +9 stride. The PCs for (2,0) and (1,2) see a stride pattern of (+2,+2,+5) -- so no consistent stride, though an offset of +9 will always be correct even from a PC-local perspective.

{On an out-of-order processor, the situation worsens: the PC that triggers the cache miss is often not the first in program order. Any of the four PCs that touch each cache line may issue its load first, and as a result, no individual PC consistently observes the +9 stride. Berti consequently underestimates the effectiveness of the +9 offset at each PC and fails to issue it, even though it would be accurate whenever used. A global offset prefetcher, such as BOP or MLOP, evaluates offsets across all PCs collectively. They therefore correctly identify all positive offsets as high-coverage since  every cache line is eventually accessed regardless of which PC reaches it first. }

\subsection{Limitations to Global Coarse-Grain Filtering}
\label{ssec:bop_limitations}

Nevertheless, BOP and MLOP are not without their own substantial limitations. BOP and MLOP tend to be overly aggressive and inaccurate in many workloads. Figure \ref{motivation:bertibopdram} illustrates that BOP and MLOP consistently exhibit a higher DRAM traffic footprint than Berti, exemplified in \bench{xz}, where BOP generates almost double the DRAM traffic compared to the baseline. In \bench{xalancbmk}, MLOP generates more than 1.5$\times$ the baseline's DRAM traffic. This is caused by the accuracy of MLOP and BOP being substantially lower than Berti in many workloads. In \bench{mcf}, MLOP achieves a substantial speedup despite low accuracy and high DRAM usage. In \bench{omnetpp} however, this inaccuracy and traffic makes the performance of MLOP worse than the baseline. This is exacerbated in bandwidth-constrained or multicore scenarios due to DRAM contention. 

To explain these shortcomings, {consider the example from \cref{fig:milc}, but where the call site accesses the array of matrices non-sequentially on each invocation. Each PC issues only a single load before execution moves elsewhere in the array, so no same-PC stride or offset can be learned.} Cross-PC offsets can still be effective, but only when prefetched at the correct load (determined by its PC). Whether a particular cache-line offset is correct is dependent on whether it falls within the following 144-byte matrix. Consider the offsets +1 and +2. For the PC loading element [0][0], the entire matrix lies ahead, so both offsets +1 and +2 would always be accurate. For the last element [2][2] neither +1 or +2 would ever be accurate. For other elements within the matrix ([0][1], [0][2], [1][1]...), the accuracy of offsets +1 and +2 become progressively lower. Moreover, simply prefetching +1 and +2 at the PC that accesses [0][0] would already cover every cache line in the matrix, making all other prefetch attempts either redundant or inaccurate.

In this paper, we show that high-performance global offset prefetching can be achieved without excessive extra DRAM traffic. We argue that the ineffectiveness of MLOP and BOP's throttling mechanisms stems from their coarse-grained offset filtering, which fails to consider the accuracy of the offsets extracted differs between PCs. Consequently, offsets accurate for only a subset of PCs disproportionately influence the scoring mechanisms of BOP and MLOP, leading to many inaccurate prefetches and unnecessary traffic. 

\subsection{Towards Better Offset Prefetching}

Collectively, these observations suggest that recent prefetching techniques exhibit limitations in either their training mechanisms or filtering strategies. Berti's approach to training local offsets inhibits it from extracting offsets that are hidden from the PC-local view. BOP's and MLOP's approach to offset filtering makes it prone to issuing excessive and inaccurate prefetches. \textbf{These observations highlight the need for offset prefetchers that simultaneously maintain broad global offset coverage while applying accuracy throttling in a local execution context.} 

\label{sec:caerus:considering_multiple_global_offsets}

\begin{figure}[t]
    \centering
    \includegraphics[width=0.9\columnwidth]{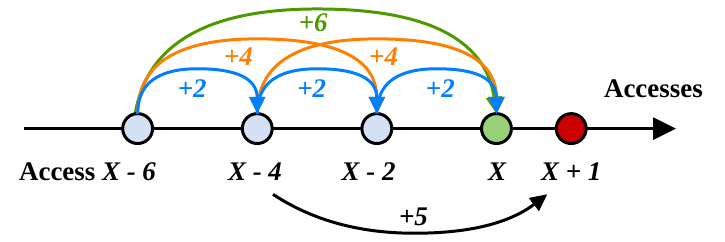}
    \caption{Naively choosing $N$ best offsets from offset training introduces large amounts of overlap: offsets of +2, +4 and +6 all cover the same accesses, leaving address $X+1$ not covered by any prefetch. Caerus avoids this by considering the next-best offset in each training round, which is the best offset for loads currently not covered by another prefetch.}
    \label{fig:overlap}
\end{figure}

An examination of existing offset prefetchers also motivates Caerus' use of multiple offsets. BOP only derives one best offset, mainly citing the increase in DRAM traffic as the reason not to consider multiple offsets simultaneously. However, as observed from Figure \ref{motivation:bertibopdram}, DRAM traffic can still become problematic when only training and issuing at most one offset per access due to poor filtering. MLOP~\cite{shakerinava_multi-lookahead_2019} issues many global offsets, varying in timeliness, but this only exacerbates DRAM contention further. Berti shows multiple offsets used in a more judicious way, by demonstrating that the best delta varies based on program context. In Berti, 16 offsets are trained simultaneously at each PC. With PC-local training and filtering, Berti is able to substantially lower DRAM-traffic overhead compared to BOP, despite training multiple offsets.

However, training multiple offsets globally introduces the challenge of \textit{prefetch overlap}. For example in \cref{fig:overlap}, assume that offset 2 covers 70\% of memory accesses -- and the remaining 30\% require offset 5. Offset 5 is very unlikely to be the second, third or fourth best offset. Instead, the next-best offsets will often be 4, 6 and 8: multiples of 2 that will prefetch more aggressively but (assuming 2 is a timely offset) add no more coverage or performance\footnote{The author of BOP~\cite{michaud_best-offset_2016} briefly acknowledges this when justifying the use of only one offset. MLOP~\cite{shakerinava_multi-lookahead_2019} attempts to solve this problem by only allowing numerically distinct offsets, however this does not solve overlapping prefetches originating from different accesses. Berti mitigates this by isolating offsets to each unique load instruction. This inhibits prefetch overlap since it prevents prefetches from one PC to cover the accesses of another PC, meaning only locally discovered offsets can cover PC-local accesses. }. 

Caerus addresses the issue of prefetch overlap by isolating the training of different offset candidates. It asks the question: assuming prefetches are issued for the offsets that have already been trained, what is the \textit{next-best offset} (\cref{ssec:nextbest}) that improves coverage the most for the remaining demand misses? 

\ignore{
\begin{table}[t]
\centering
\setcellgapes{3pt} %
\makegapedcells 
{
\begin{tabular}{|c|c|}
\hline
\makecell{\textbf{Limitation of} \\ \textbf{Current Proposals}} & \textbf{Solutions in Caerus} \\
\hline
\makecell{Low Coverage due to \\ \textbf{PC-local Fine-Grain Training} \\ (Berti)} & \makecell{Deriving Offsets from a Global Scope \\ (\cref{ssec:offsets})}  \\
\hline
\makecell{\textbf{Overlapping Prefetches} \\ (BOP, MLOP)} & \makecell{Overlap Prevention \\ (\cref{ssec:overlap})}  \\
\hline
\makecell{High DRAM Traffic due to \\ \textbf{Coarse Grain Filtering} \\ (BOP, MLOP)} & \makecell{Per-PC Filtering on Global Offsets \\ (\cref{ssec:accuracytracking})}  \\
\hline
\end{tabular}
}
\caption{{Limitations in offset prefetchers solved by Caerus.}}
\label{tab:limitation_to_solution}
\end{table}
}

\section{Caerus}\label{sec:caerus}

\begin{figure*}[t]
    \centering
    \includegraphics[width=\linewidth]{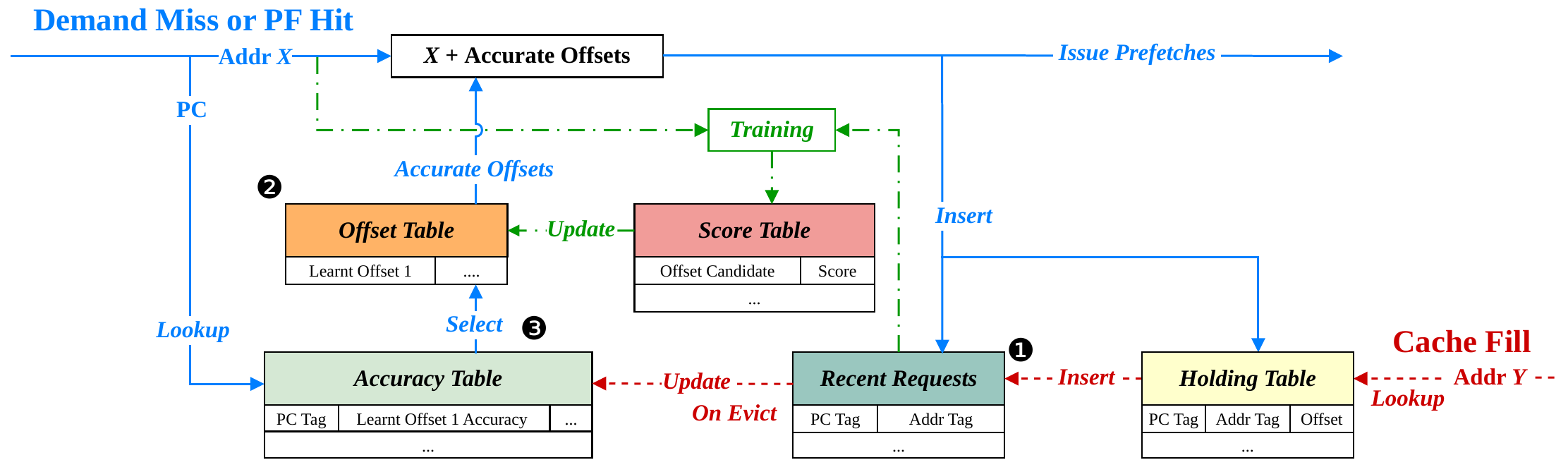}
    
    \caption{\revision{Design of Caerus}. On a demand miss/prefetch tag hit (\textcolor{fig6blue}{solid blue}), Caerus generates accurate offsets for prefetching (\cref{ssec:accuracytracking}). The accessed address is inserted into the Holding Table or Recent Requests (RR) Table (\cref{ssec:history}). The Score Table is updated if deltas between the access and entries in the RR Table match an offset candidate (\textcolor{fig6green}{dot-dashed green}) (\cref{ssec:offsets}). Upon a cache fill (\textcolor{fig6red}{dashed red}), matching Holding-Table entries update the RR Table (\cref{ssec:history,ssec:accuracytracking}). Illustrations are given for \ding{182} RR-Table operations (\cref{caerus:rr_table}), \ding{183} offset retraining (\cref{caerus:retrain}), and \ding{184} accurate offset selection (\cref{caerus:accuracy}). } 
    \label{caerus:organisation}
\end{figure*}

To address the limitations of contemporary offset prefetchers, here we present Caerus: an offset prefetcher that employs PC-local filtering on overlap-resistant globally derived offsets.  \revision{
To cover the diverse access patterns across multiple PCs, Caerus employs a novel circular training mechanism that learns multiple \textit{non-overlapping} offsets. Caerus uses the \textbf{Score Table} to evaluate offset candidates against recent global accesses in the \textbf{RR Table}. The learned offsets are committed to an \textbf{Offset Table} and periodically \textit{retrained}. Using the \textbf{Holding Table}, Caerus is able to ensure timeliness of its learned offsets by inserting entries into the RR Table after its fill delay. Caerus then filters the derived offsets through its per-PC \textbf{Accuracy Table}, allowing it to adapt to different PC contexts and facilitate cross-PC prefetching. 
}

 \revision{\Cref{caerus:organisation} shows the design of Caerus using the following structures, with \Cref{alg:caerus:pseudocode_for_caerus} showing an operational timeline}. 
\revision{
\begin{itemize}[leftmargin=*]
     \item \textbf{Score Table}: records scores for offset candidates during the training phase (\cref{ssec:offsets}).
     \item \textbf{Offset Table}: tracks learned active offsets (\cref{ssec:offsets}).
    \item \textbf{Recent Requests (RR) Table}: a set of shift registers that store the PC and address tag of recent prefetches or demand misses. Entries in the RR Table are used for offset training (\cref{ssec:offsets}) on new observations and accuracy tracking  (\cref{ssec:accuracytracking}) upon eviction. 
    \item \textbf{Accuracy Table}: a PC-indexed table containing a 4-bit saturating counter for each active offset to track its accuracy (\cref{ssec:accuracytracking}).
    \item \textbf{Holding Table}: tracks the PC and address tag of in-flight prefetch requests/demand misses to maintain timeliness (\cref{ssec:history}). 
\end{itemize}
}

\setlength{\fboxsep}{6pt}
\begin{figure}
\small
\fbox{%
\begin{minipage}{\dimexpr\linewidth-2\fboxsep-2\fboxrule\relax}
\begin{algorithmic}
\Statex \textbf{On demand miss or prefetch hit at addr $X$ at PC $A$:}
    \State $Ps$ = $X$ + AccTable.getAccOffsets(PC $A$) 
    \State Issue $Ps$ $\to$ PrefetchQueue
    \If {len($Ps$) $>$ 0} 
        \State HoldingTable.push(PC $A$, addr $Ps[0]$\protect\footnotemark, Offset $Ps[0] - X$)
    \ElsIf {Demand Miss}
        \State HoldingTable.push(PC $A$, addr $X$, Offset $0$)
    \Else 
        \State AccTable.updateAcc(RRTable.getVictim(addr $X$))  
        \State RRTable.push(PC $A$, addr $X$)
    \EndIf
    \State offsetTraining(addr $X$) 
\\

\Statex \textbf{On cache fill at addr $Y$:}
    \If {HoldingTable.contains(addr $Y$)}
        \State (PC $C$, addr $Z$, Offset $D$) = HoldingTable.pop(addr $Y$)
        \State AccTable.updateAcc(RRTable.getVictim(addr $Z-D$))  
        \State RRTable.push(PC $C$, addr $Z$ - $D$)
    \EndIf
\end{algorithmic}
\end{minipage}}
\caption{\revision{Operational timeline of Caerus. }}
\label{alg:caerus:pseudocode_for_caerus}
\end{figure}

\subsection{Offset Training}\label{ssec:offsets}
\revision{
Caerus employs a novel training mechanism that builds upon BOP's, Berti's and Chimera's training mechanisms to learn multiple \textit{overlap-resistant} \textit{next-best} offsets. 
}

The original BOP maintains a list of predetermined offset candidates and cycles through each offset candidate in turn, testing them in its Recent Requests table: a direct-mapped array of the last few memory accesses that could have generated timely prefetches (see \cref{ssec:history}). This is tested on every demand miss or tagged prefetch hit (i.e., the first time a previously prefetched value is accessed from the CPU-side, which is when a demand miss would otherwise have occurred without the prefetcher). When a cache line with address $X$ is accessed
\footnotetext{\revision{$Ps[0]$ is the first prefetch that did not hit in the cache or holding table. }}
\footnote{Addresses are cache-line aligned; offsets are expressed in numbers of cache lines.}, BOP tests for $X - D$ in the Recent Requests (RR) table. If it is present in the RR, $D$'s score is incremented. The learning phase ends when an offset’s score reaches the \texttt{ScoreMax} threshold, or when the number of rounds (the number of times all of BOP's offsets have been tested) reaches the \texttt{RoundMax} threshold. 

This is a slow process and we observe that BOP typically requires around 5000 memory accesses to train a reliable offset. This is because every offset is considered one at a time for each access, even if they never appear in the application. Berti uses a structure that reverses this process to allow for faster training. Rather than walking through a list of offsets and testing for their existence, it instead stores a table of recent accesses indexed by PC. Offsets are then generated by taking, for each historical access $A$ at the same PC that is deemed able to generate timely prefetches, generating $D$ = $X-A$ instead of testing for $A = X-D$ for each offset $D$. The intuition is that, by using a heuristic that good offsets are localized to same-PC accesses, fewer candidates need to be checked. 

\begin{figure}
    \centering
    \includegraphics[width=\linewidth]{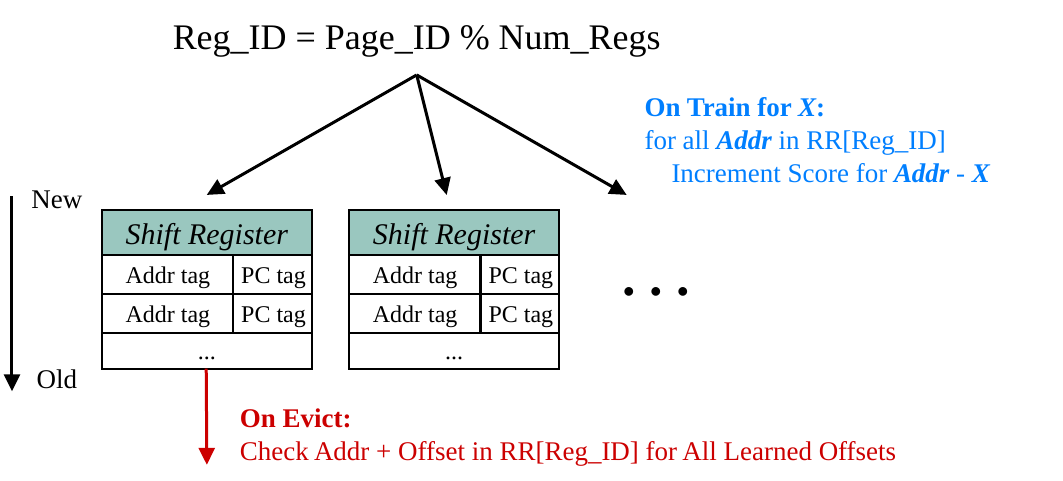}
    
    \caption{Illustration of the operations on the Recent Requests (RR) table. When training a new address X, the scores of the deltas between X and the entries in the shift register with the same Reg\_ID are incremented. When a shift register is full, each new insertion evicts the oldest entry (FIFO) and updates the accuracy of the learned offsets. }
    \label{caerus:rr_table}
\end{figure}

\revision{Caerus uses a similar heuristic: localizing to testing (though not exclusively issuing\footnote{\revision{Since Caerus operates on virtual addresses, it issues cross-page prefetches in the same manner as Berti~\cite{navarro-torres_berti_2022}, even though its offsets are trained within a single page. In effect, each instance of a cross-page offset is conservatively assumed to be inaccurate for the purposes of training offsets and for per-PC filtering (\cref{ssec:accuracytracking}). Consider a consistently correct offset of +1:  63 out of every 64 cache lines in a 4KiB page will be in the same page, and so the prefetch will reach threshold for issuing including cross-page. We found that increasing the tracked region size to 8KiB and 16KiB only brings 0.46\% and 1.92\% more \textit{cross-page} prefetches, and no more speedup, as Caerus already issues them effectively.}})  same-page accesses}\footnote{Chimera~\cite{he_chimera_2024} uses a similar page-localization heuristic, but trains PC-local offsets like Berti and region-local offsets indexed by region. We found that region-local offsets took too long to train, meaning we had left the region by the time they were useful.}. Our modified RR table is thus stored as a set of 32 shift registers, each storing 16 entries (see \cref{caerus:rr_table}). The relevant shift register is chosen based on the page index. Within the shift register, all accesses within the same page (measured by both index and tag) as the current access \textit{X} are compared to generate offsets. \revision{Any generated offsets that are also in the BOP-style offset candidate list\footnote{Here we use the same list as BOP~\cite{michaud_best-offset_2016}: positive and negative offsets of magnitude less than 64 and of the form $2^i\times3^j\times5^k$, with $i,j,k\geq0$.} are incremented. At the end of a 500-load epoch, or when one entry reaches a \texttt{ScoreMax} threshold of 24\footnote{\revision{Both are proposals from BOP but with adjusted parameter sizing to reflect faster training.}}, the best offset is chosen and committed to the Offset Table so it can issue prefetches. Similar to BOP's \texttt{BadScore} mechanism, we disable the current learning offset slot if no offset in the Score Table scores above 10.}

The shift registers act as FIFO queues: once full, every insertion results in the oldest entry being evicted from the RR. This evicted entry is used alongside currently learned offsets to update the accuracy of PC-offset pairs in Caerus' Accuracy Table (\cref{ssec:accuracytracking}). 

\subsubsection{Discovering the Next-Best Offset} \label{ssec:nextbest} {We alter BOP's training mechanism in Caerus to train \textit{N} best offsets}. The naive method of taking the \textit{N} best scoring offsets in the same learning phase is inadequate, since the selected offsets are prone to overlap (see \cref{fig:overlap}), harming the prefetcher's potential to discover groups of offsets that collectively maximize coverage. 

As a result, after selecting the best-scoring offset from the last learning phase, we want to discover the \textit{next}-best offset, i.e., the offset that increases prefetch coverage the most in the presence of other (typically better) offsets. We alter BOP's offset training to only occur either on a cache miss (where there were no timely prefetches by any other offset), or when we are performing \textit{cyclical retraining} (\cref{ssec:cyclical}) on the offset that caused the prefetch for the current access. Unlike BOP, we do not train offsets on any other tagged-prefetch hits already inside the cache\ignore{Though we do train on tagged-prefetch MSHR hits, which represent late prefetches that are not currently covered by timely means.}. This means that each training round sees a filtered address stream: rather than trying to find the overall best offset each time, we instead attempt to find the best offset for all the accesses not covered by prefetches already being triggered existing learned offsets (\cref{ssec:overlap}). 

\begin{figure}
    \centering
    \includegraphics[width=\linewidth]{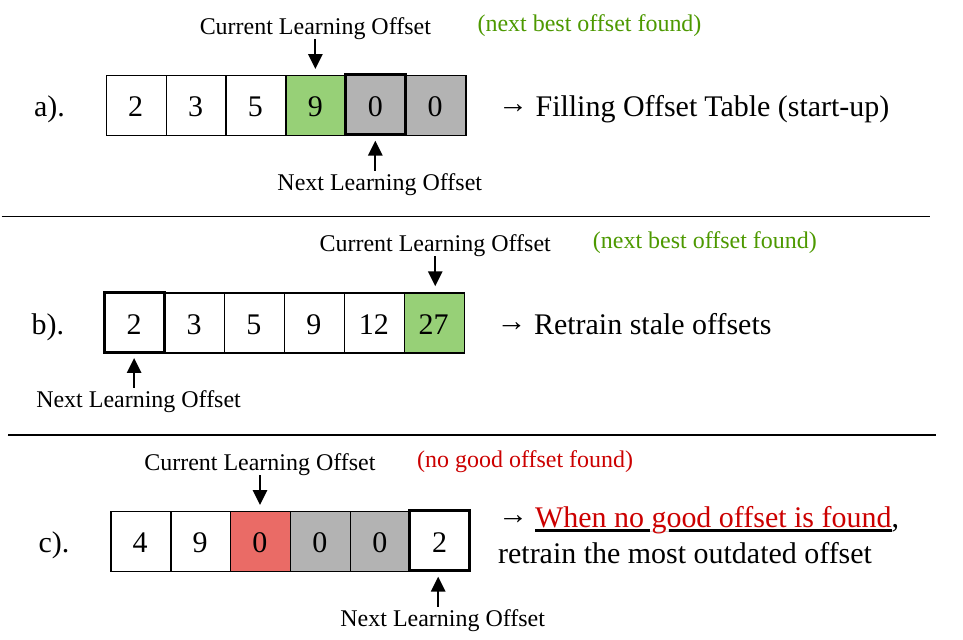}
    
    \caption{Illustration of cyclical retraining. The Offset List is first filled with next best offset candidates (see \cref{ssec:nextbest}). Caerus then continuously retrains its most outdated offsets to reflect changes in the memory access pattern. }
    \label{caerus:retrain}
\end{figure}

\subsubsection{Cyclical retraining} \label{ssec:cyclical} 

\revision{Caerus continually adapts to program phase changes by replacing outdated offsets in the Offset Table. On start-up, the Offset Table is progressively filled with newly trained next-best offsets (\cref{caerus:retrain}\textit{.a}). Once the table is full, Caerus begins \textit{retraining} the oldest slot: it runs a new learning phase for that slot to determine whether a better offset exists (\cref{caerus:retrain}\textit{.b}). Unlike initial training, which only considers accesses not already covered by any current offset (\cref{ssec:nextbest}), retraining also considers accesses covered by the offset currently occupying the slot. Hence, the incumbent offset competes alongside new candidates, allowing it to be confirmed or replaced. If the learning phase fails to find any good offset, the slot is disabled and Caerus cyclically scans the table, skipping disabled slots, to locate the next oldest active slot to retrain (\cref{caerus:retrain}\textit{.c}). Disabled slots are naturally re-enabled: after a successful learning phase, the learning pointer advances to the next slot in round-robin order \textit{without} skipping disabled slots, so a previously disabled slot is eventually reached, trained from scratch, and re-populated with a new offset should one exist. }

In contrast to offset training, which aims to train on accesses not covered by prefetches from any current offset (\cref{ssec:nextbest}), offset \textit{re}training allows accesses covered by the current learning offset to be considered as well. This allows the current retraining offset to be unchanged if no better candidate is found. Otherwise, when offset O at position N is retrained with a new non-overlapping offset P, position N+1 ends up being retrained back to the original offset O. As a result, offset $O$ is temporarily removed from its original position and remains unavailable until it is rediscovered in a later training epoch, which can reduce coverage.

\subsection{Overlap Prevention} \label{ssec:overlap}

Caerus aims to train only on accesses not covered by other offsets to prevent prefetch overlap from consuming the finite supply of trained offsets (see \cref{ssec:offsets,fig:overlap}). This is achieved by filtering out loads that would be covered by prefetches from existing active offsets. Upon observation of an access to an address $X$ from PC $A$, where $R$ is the currently training offset, $X$ is only trained on if there is no {currently learned} offset $D$ for $D\neq R$ such that: \textit{a)} could have prefetched $X$ ($X-D$ is in the RR table), and \textit{b)} is also accurate (PC $A$ is issuing $D$ as an offset). 

{If such an offset $D$ exists, and PC $A$'s accuracy table for $D$ passes the accuracy threshold, then Caerus is currently using offset $D$ to issue prefetches. Since the relevant address $X-D$ is in the RR table, the resulting prefetch would also be timely, so no other offset exists that could improve Caerus' ability to issue timely prefetches for address $X$. By contrast, if no offset $D$ satisfies the above criteria, that means the current access $X$ is either only covered by timely prefetching with offset $R$, or is not covered at all. In both cases, $R$ should be trained to attempt to cover the access $X$ in the future}. 

\subsection{Accuracy Tracking}\label{ssec:accuracytracking}

The Accuracy Table is indexed by the PC and each entry holds a 4-bit saturating-counter value for each active offset\footnote{Per-PC filtering of offsets breaks BOP's accuracy-coverage coupling, where in BOP the highest-coverage offset is also the most accurate (as offsets may be highly accurate at some PCs but not others, thus low coverage but high accuracy). The filtered address stream seen by the next-best training algorithm (\cref{ssec:nextbest}) also breaks this link. Collectively, both decisions mean that in Caerus, accuracy is treated independently from choosing the best offset.}. 

Accuracy of offsets is updated upon each eviction from the Recent Requests (RR) table, rather than on insertion, to see {whether any future} accesses would be successfully prefetched by an offset $D$ from the perspective of the evicted address $E$. A set of candidate prefetch addresses is derived by summing the current offsets ($D_1$, $D_2$, …) with address $E$.  If $E + D_n$ is found in the corresponding shift register of the RR that $E$ was just evicted from, $D_n$'s accuracy for the PC that initiated $E$ is incremented. 
Conversely, if $E + D_n$ was not found its accuracy is decremented. We find superior performance by also supporting a notion of \textit{indifference}: if the distance between $E$ and $D_n$ in the shift register is greater than 2/3 the length of the shift register, we consider this to be a prefetch that is likely to be used within a distance where it will be an L2 hit but not an L1D hit, and so we neither increment nor decrement the counter to represent a situation where we are \textit{indifferent} to prefetching. This avoids over-pessimism and overoptimism in fetching.

\begin{figure}
    \centering
    \includegraphics[width=\linewidth]{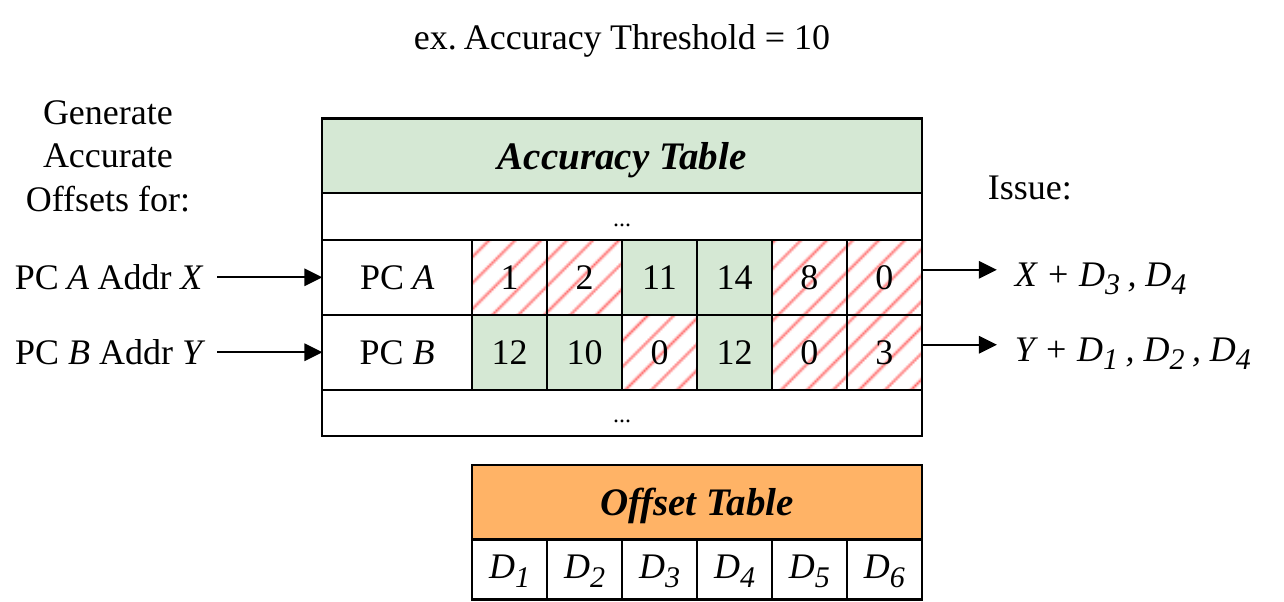}
    
    \caption{Illustrative example of generating accurate prefetches. The PC of the demand miss/prefetch tag-hit is looked up in the accuracy table, and currently active offsets that have a sat-counter value larger than the accuracy threshold generate prefetches. If no offsets have a sat counter value that reaches the threshold, no prefetches are generated.}
    \label{caerus:accuracy}
\end{figure}

On every cache miss or prefetch hit in the cache, the PC that generated the demand request is queried in the accuracy table. Offsets with an accuracy exceeding the preset accuracy threshold {in the relevant saturating counter} are gathered, summed with the demand address, and issued into the prefetch queue (see \cref{caerus:accuracy})\footnote{{To issue a prefetch, only the offset table and accuracy table are queried. Operations on the other three tables are pipelined, so they do not affect prefetch issue latency.}}. 

\subsection{Managing History and Timeliness}\label{ssec:history}

Like BOP, Caerus inserts entries into the Recent Requests (RR) table with a delay to ensure the timeliness of its trained offsets. For a prefetch request for $X+D$ at demand access at $X$ to be timely, it should arrive before the demand access of $X+D$. Hence, for a prefetched request or demand miss, the source address $X$ is moved into the RR when the requested prefetch $X+D$ or demand miss $X$ is filled in the L1D cache. When a subsequent access with address $Y$ is trained against address $X$ in the RR table, the offset $Y-X$ would have been a timely prefetch\footnote{In BOP and Caerus, the simplifying assumption is made that any prefetch issued from $X$ would be of similar timeliness. This mechanism aims to find the distribution of latencies rather than latency of each request~\cite{michaud_best-offset_2016}. } issued from the demand access for $X$.

In BOP, in any given training interval, either no prefetches are issued (in which case every demand miss can be moved into the RR table at the time of its own fill), or exactly one prefetch is issued per demand miss or prefetch-tag hit with the same offset $D$. It thus always serves, on the fill of line $X$, to place $X-D$ into the RR table (with $D=0$ representing when prefetching is turned off).

For Caerus the answer is more complicated. Within a training interval, a demand miss or a prefetch-tag hit, {i.e., a line that is a cache hit only because it was a useful prefetch,} may trigger 0, 1 or multiple prefetches at several possible offsets (depending on their per-PC accuracy). {To track when the observed tag-hit to a line $X$ should cause $X$ to be inserted into the RR table (which should only occur when $X$ is able to fulfill a timely prefetch request), at least one of these prefetches must be tracked. Once the fill for that prefetch occurs, $X$ is inserted into the RR. This access must also be a prefetch to an address that does not hit in the MSHR, otherwise we would not observe the full latency of a prefetch issued at the time of $X$'s observation.} This requires the addition of a new structure: for the target in-flight request, the Holding Table stores the base address of the request, the offset of the prefetch (or 0 for demand misses), and the PC for accuracy tracking once the value is evicted from the RR (\cref{ssec:accuracytracking}). 

Since Caerus attempts to capture cross-PC accesses, untimely prefetches may be useful due to the involved fine-grained dependencies preventing timely alternatives. We use an \textit{early insertion} heuristic: if on an access to address A, no entries within the same page as A are inside the RR table, A is moved immediately into the RR table to eagerly train offsets even if they may be untimely.

Hence, every observation (miss or prefetch-tag hit) ultimately results in an entry being placed in the RR table. There are four possible scenarios for an observation:

\begin{itemize}[leftmargin=*]
\item Address $X$ is the first access in a page. $[PC,X]$ is immediately inserted into the RR table. 
    \item Address $X$ generates one or more non-redundant prefetches. One such prefetch, $X+D_i$, is chosen as representative, and $[PC,X+D_i,D_i]$ is placed in the Holding Table to insert $[PC,X]$ into the RR table once the prefetch request is filled. 
    \item Address $X$ generates no prefetches but is itself a cache miss. $[PC,X,0]$ is placed in the Holding Table so that X is inserted into the RR once its access latency completes. 
    \item Address $X$ generates no prefetches but is itself a prefetch tag hit. No latency can be measured, so $[PC,X]$ is immediately inserted into the RR table. 
\end{itemize}

\renewcommand{\arraystretch}{1.5} 
\begin{table}
\begin{center}
    \begin{tabular}{|p{1.2cm}|p{6.5cm}|}
    \hline  
\textbf{Core} & Out-of-order, {TAGE-SC-L} Branch Predictor, 4 GHz with 6-issue width,
5-retire width, 352-entry ROB \\ \hline
\textbf{TLBs} & L1 iTLB \& dTLB:  64 entries, 4-way, 1 cycle 

STLB: 1536 entries, 12-way, 8 cycles \\\hline
\textbf{L1I} & 32 KiB, 8-way, 3 cycles \\\hline
\textbf{L1D} & 48 KiB, 12-way, 4 cycles, LRU, non-inclusive \\ \hline
\textbf{L2C} & {2 MiB} 16-way, 10 cycles, LRU, non-inclusive \\\hline
\textbf{LLC} & {6 MiB / core}, 16-way, 20 cycles, LRU, non-inclusive \\\hline
\textbf{MSHRs} & 8/16/32/64 in L1I/L1D/L2/LLC \\\hline
\textbf{PQs} & 8/8/16/32 in L1I/L1D/L2/LLC \\\hline
\textbf{DRAM} &
4GB, 3200MTPS, 8 banks, 4 KiB row-buffer per bank
\\ \hline
\end{tabular}
\end{center}
\caption{Simulation parameters for baseline system.}
\label{tab:core}
\end{table}

\section{Experimental Setup}

Similar to other recent designs~\cite{navarro-torres_berti_2022,he_chimera_2024,Hyperion}, we evaluate Caerus using the trace-based simulator ChampSim~\cite{gober2022championshipsimulatorarchitecturalsimulation} using the configuration shown in \Cref{tab:core}. This paper selects a wide range of offset prefetchers to evaluate the performance of Caerus. We compare against a carefully-tuned degree-8 stride prefetcher. We also compare against the state-of-the-art offset prefetcher Berti~\cite{berto}, using its open source implementation. Other competitors include Chimera~\cite{he_chimera_2024}, provided by the original authors, the open-source implementation of MLOP~\cite{shakerinava_multi-lookahead_2019}, and our own implementation of the Best Offset Prefetcher~\cite{michaud_best-offset_2016} based on gem5's implementation. Their configuration and sizing is  provided in \cref{tab:eval_config}. We also evaluate GBerti~\cite{Posluns2026GlobalBerti} and BertiGO~\cite{Singh2026BertiGO} from DPC4~\cite{dpc4_2026} in their default configurations, at 12.17 KiB and 29.72 KiB respectively. All prefetchers evaluated are placed in the L1D cache\footnote{BOP was evaluated as an L2 prefetcher in its original proposal~\cite{michaud_best-offset_2016}. We evaluate BOP in the L1D since it has similar performance to BOP in the L2, and for purposes of comparing like-for-like coverage and accuracy with other L1D prefetchers.}, observe virtual addresses and issue virtual prefetches that trigger page-table walks. A multi-level prefetching study with both L1D and L2 prefetchers is also conducted, shown in \cref{ssec:l2}.

\begin{table}[t]
\begin{center}
    \begin{tabular}{|c|c|c|}
    \hline
 \textbf{Tables} & \textbf{Entries} & \textbf{Size} \\
 \hline
\textbf{Accuracy} & 128 $\times$ 4b Sat Counter $\times$ 20 Offsets & 1280B \\
\textbf{Holding} & 64$\times$(7b PC Tag, 14b Addr Tag, 7b Offset)& 224B \\
\textbf{RR} & 32 $\times$ 16 $\times$ (14b Addr Tag, 7b PC Tag) & 1344B \\
\textbf{Score} & 104 $\times$ 5b Score & 65B \\
\textbf{Offset} & 20 $\times$ 7b Offset & 18B \\
 \hline
{\textbf{Total}} & & 2.86KiB  \\\hline
\end{tabular}
\end{center}
\caption{Sizing of Caerus' structures.}
\label{tab:sizetable}
\end{table}

\begin{table}[t]
\begin{center}
    \begin{tabular}{|p{2.3cm}|p{4cm}|p{1cm}|}
    \hline
\textbf{Stride}~\cite{stride} & 256 sets, 4 ways, degree 8 & 12.0KiB \\ \hline
\textbf{BOP}~\cite{michaud_best-offset_2016} & 256-entry RR, 104 offset candidates & 0.44KiB \\ \hline
\textbf{MLOP}~\cite{shakerinava_multi-lookahead_2019} & 128-entry AMT, 500 update epoch, degree 16 & 6.56KiB \\ \hline
\textbf{Chimera}~\cite{he_chimera_2024} & 128-entry HT, 16-entry PP, 16-entry SP, 1-entry EP & 2.36KiB\\ \hline
\textbf{Berti}~\cite{navarro-torres_berti_2022, berto} &8-set 16-way history table, 16-entries 16-deltas delta table & 2.55KiB \\ \hline
\end{tabular}
\end{center}
\caption{Configuration of comparison prefetchers.}
\label{tab:eval_config}
\end{table}

In total, we evaluate on 307 traces of simpoints across four benchmark suites, originally provided by the 3rd Data Prefetching Championship ~\cite{champsim_traces_speccpu_2025} and open-sourced by Pythia ~\cite{Pythia}. These include simpoints from SPECspeed 2017 (95 traces), SPEC CPU2006 (93 traces), Ligra (76 traces), and PARSEC 2.1 (43 traces). Each trace is warmed up for 50 million instructions before being timed for a 200 million instruction window. All presented results are aggregated based on the combined simpoints for each workload, and weighted whenever weights are available.

\Cref{tab:sizetable} gives the sizing of Caerus' structures, totaling 2.86 KiB. In the evaluated configuration of Caerus, the RR table consists of 32 shift registers with 16 entries each. The accuracy table contains 128 entries, indexed using 7 bits. For each address, we store a 14-bit hash as an address tag. Potential offsets are selected from a list of 104 candidates, each assigned a 5-bit score. Compared to other state-of-the-art prefetchers, Caerus is only 0.31 KiB larger than Berti and 3.70 KiB smaller than MLOP (see~\cref{tab:eval_config}).

\begin{figure*}[t]
\resizebox{\linewidth}{!}{%
    \includegraphics{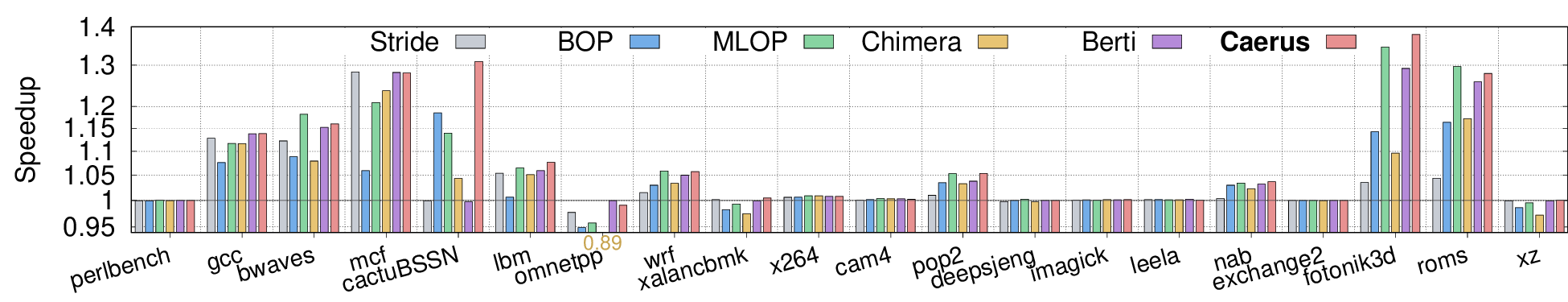}
    \includegraphics{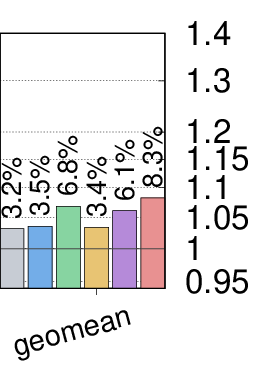}
}
    \caption{Speedup on SPECspeed 2017: Caerus outperforms other techniques overall by succeeding in both PC-local and global prefetching settings, while avoiding contention from excess DRAM traffic harming performance.}
    \label{fig:speedup17}
    \label{fig:speedup}
\end{figure*}

\section{Evaluation}

In SPECspeed 2017, Caerus achieves an 8.3\% speedup over a baseline with no prefetching, where a well-tuned stride prefetcher achieves 3.2\%. Caerus' closest performing competitor, MLOP, achieves only 6.8\%, with 20\% excess DRAM traffic versus Caerus' 4.7\%. Berti, due to its PC-local context, achieves only 6.1\%, and recent DPC4 extensions BertiGO and GBerti do not meaningfully improve this (6.6\% and 5.8\%) despite an order-of-magnitude increase in hardware overhead. Chimera and BOP achieve considerably lower speedup because of high DRAM traffic overheads and low coverage respectively. Results in other workloads follow a similar trend. Caerus also performs well as an L2 prefetcher. When deployed at both cache levels, it outperforms every other combination of state-of-the-art L1D and L2 prefetchers while incurring extremely low traffic and resource overheads.

\begin{figure}[t]
    \centering
    \includegraphics[width=\columnwidth]{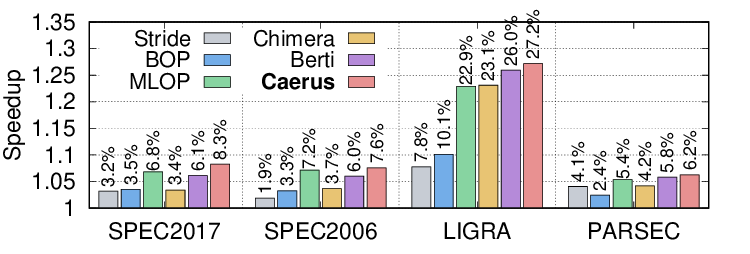}
    \caption{Speedup of evaluated prefetchers across suites.}
    \label{fig:benches-ipc}
\end{figure}

\begin{figure}[t]
    \centering
    \includegraphics[width=\columnwidth]{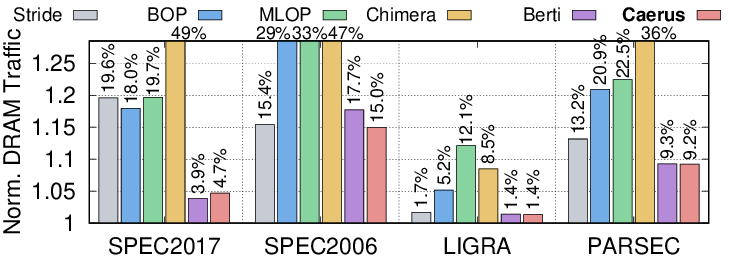}
    \caption{Normalized DRAM traffic of evaluated prefetchers.}
    \label{fig:benches-dram}
\end{figure}

\subsection{Single-Core Performance Analysis}

\Cref{fig:speedup} shows a breakdown of speedups per-workload in SPECspeed 2017. Caerus tends to outperform rival techniques by being highly consistent in achieving speedups across different types of memory access. One notable example is \bench{cactuBSSN}, where Caerus is the most performant prefetcher by a large margin. Although loads in \bench{cactuBSSN} technically follow regular stride patterns, Berti is not able to provide any speedup due to the overwhelming number of these interleaved loads. Global prefetchers like BOP and MLOP are able to capture some global offsets, but their performance is limited due to finding single or redundant offsets that overlap with each other. Caerus is able to achieve the best of both worlds by finding numerous overlap-resistant offsets while efficiently determining their PC-local usefulness. Other workloads such as \bench{pop2} and \bench{fotonik3d} feature deltas that are simple for MLOP and Caerus to pick up cross-PC but are hard for Berti to pick up locally. Caerus is only rarely not the best performing workload: in \bench{bwaves} and \bench{roms}, MLOP achieves marginally higher speedup than Caerus. MLOP's aggression on specifically subtimely prefetching gives a slight gain.

In most workloads, Chimera underperforms due to heavy aggression along with a lack of explicit timeliness consideration in its design unlike the other techniques. BOP suffers from low coverage and high traffic due to its design. 

\Cref{fig:benches-ipc} show that other workloads follow the same trends. In SPEC CPU2006, workloads such as \bench{milc} (\cref{fig:milc}) benefit from complex pattern tracking. In \bench{milc}, Caerus was able to outperform BOP, MLOP, and Berti by 17.8\%, 12.0\%, and 1.4\% respectively. In Ligra, Caerus is able to exceed Berti's performance even on accesses that often feature solely PC-local deltas, where MLOP and Chimera incorrectly fetch data on irregular accesses. In PARSEC, Caerus is able to avoid the performance drop resulting from the high DRAM traffic of BOP and MLOP  (\bench{canneal}) while outperforming Berti in workloads such as \bench{streamcluster} due to higher coverage. 

\begin{figure}[t]
    \centering

    \includegraphics[width=\linewidth]{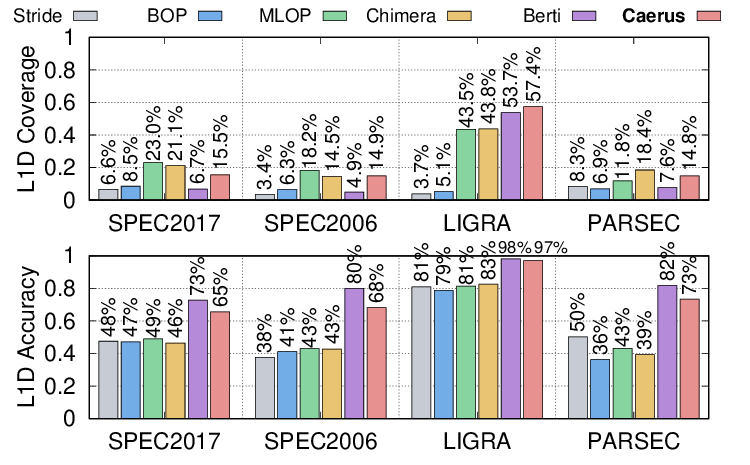}
    \includegraphics[width=\linewidth]{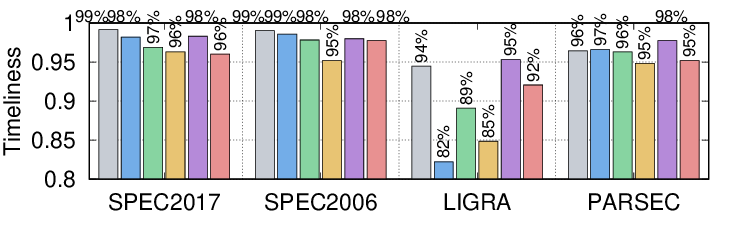}
    
    \caption{Accuracy, Coverage \revision{and Timeliness}.}
    \label{fig:benches-acccov}
\end{figure}

\subsection{DRAM Traffic}

Caerus is considerably more DRAM-efficient than other techniques (\Cref{fig:benches-dram}). Chimera has extremely high traffic due to combining BOP-like and Berti-like strategies. BOP has high traffic due to having no PC-level filtering, which is exacerbated further in MLOP which fetches up to 16 unfiltered global offsets on each observation instead of just one like BOP.

In many workloads such as \bench{xalancbmk} and \bench{xz} in SPECspeed 2017 and \bench{omnetpp} and \bench{sjeng} in SPEC CPU2006, Caerus is the only prefetcher to not under-perform the baseline. This is due to Caerus' accuracy tracking mechanism correctly identifying that prefetching has limited value in those workloads. Taking \bench{xz} as an example, Caerus incurs only 0.76\% extra DRAM traffic compared to Berti's 5.57\%, BOP's 75.6\% and MLOP's 61.6\%.

At its default tuning Caerus' DRAM traffic is marginally higher than Berti's in SPECspeed 2017, but Caerus can be retuned to achieve lower DRAM traffic while still achieving higher performance than Berti as shown in \Cref{ssec:sensitivity}. However the opposite is not true for Berti, as BertiGO is tuned to be more aggressive than Caerus but yet achieves less performance (\cref{ssec:dpc4}).

\subsection{Accuracy, Coverage, \revision{and Timeliness}}

\Cref{fig:benches-acccov} shows L1D accuracy\footnote{Accuracy is calculated as Useful Prefetches / (Useful Prefetches + Useless Prefetches), where useless prefetches refer to prefetch requests that are not used before eviction.}, coverage\footnote{Coverage is calculated as Useful Prefetches / (Useful Prefetches + L1D Cache Misses).}, \revision{and timeliness\footnote{Timeliness is calculated as Timely Prefetches / Useful Prefetches. }} for each evaluated technique across benchmark suites. 

Caerus outperforms Berti, BOP and Stride prefetchers due to significantly higher coverage as a result of being able to issue large numbers of cross-PC and same-PC offset prefetches. Compared to Chimera and MLOP, Caerus' coverage is reduced in order to improve accuracy and lower DRAM traffic. In Ligra, Caerus is able to achieve higher coverage than MLOP and Chimera due to its resistance to offset overlap.

Although Caerus' L1D accuracy is slightly lower than Berti's, this does not translate to a proportionate increase in DRAM traffic from Caerus. This is due to Caerus encouraging more aggressive movement of data between cache levels \revision{(i.e., L3 to L1D and L2). It often brings in data that is used before eviction from the on-core memory system, but not before eviction from the L1D, particularly in workloads such as cactuBSSN\footnote{\revision{In cactuBSSN, Caerus sees 5.7\% and 7.7\% more L2C accesses compared to MLOP and Berti respectively while incurring fewer L2C misses.} }.} Caerus does not directly aim for high L1D accuracy, rather it aims for high hit-rate in its recent requests table, which can exceed L1D capacity and thus make use of prefetches that, while too early to be used before eviction from the L1D, are still useful in the L2. \revision{Bringing} this data into the L1D is a conscious design decision since it enables Caerus to track request timeliness (\cref{ssec:history}) while incurring negligible performance overhead (hence Caerus is \textit{indifferent}, as discussed in \Cref{ssec:accuracytracking}).

Caerus sustains timeliness comparable to BOP even with the challenges of handling multiple filtered offsets (\cref{ssec:history}), and rises well above BOP in Ligra due to Caerus' rapid response to changes in program behavior. Caerus also achieves timeliness comparable to Berti while avoiding the L1D modifications (timestamped cache tags) that Berti requires. %

\subsection{Comparison with DPC4 Prefetchers}\label{ssec:dpc4}

\begin{figure}[t]
    \centering
    \includegraphics[width=\columnwidth]{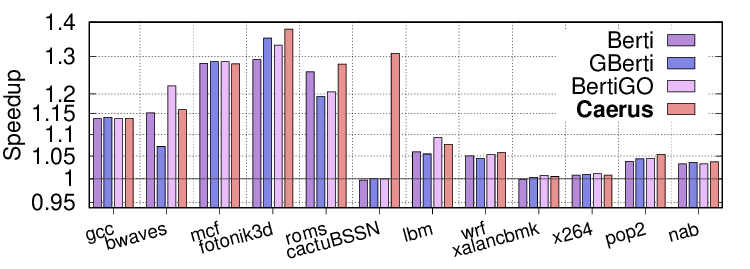}
    \caption{Speedup of Caerus, Berti, GBerti, and BertiGO in SPECspeed 2017 workloads that show a performance delta of larger than 1\%. Caerus outperforms BertiGO, the highest-performance, largest, most aggressive Berti variant, by 1.5\% in the full SPECspeed 2017 suite (\cref{fig:speedup-combined}) and up to 30\% in individual workloads.}
    \label{fig:dpc4}
\end{figure}

\Cref{fig:speedup-combined} previously showed the performance of larger extensions of Berti, the DPC4~\cite{dpc4_2026} winners BertiGO~\cite{Singh2026BertiGO} and second-place GBerti~\cite{Posluns2026GlobalBerti} against their incurred DRAM traffic and required hardware overhead. Despite requiring significantly less state, Caerus is able to outperform these two prefetchers by 1.5\% and 2.4\% respectively while incurring less DRAM traffic. \Cref{fig:dpc4} gives more insight on this performance gap. Although BertiGO's larger history enables it to beat Caerus in workloads such as \bench{bwaves} and \bench{lbm}, it lags behind Caerus in most benchmarks, especially in \bench{cactuBSSN} due to its inability to discover global offsets. GBerti attempts to identify global spatial patterns via column accesses, however its restrictive definition of global offsets hinders it from discovering effective offsets in benchmarks such as \bench{cactuBSSN} and \bench{fotonik3d}. 

\subsection{\revision{Comparison with Composite L1D Prefetchers}}

\begin{figure}[t]
    \centering
    \includegraphics[width=\columnwidth]{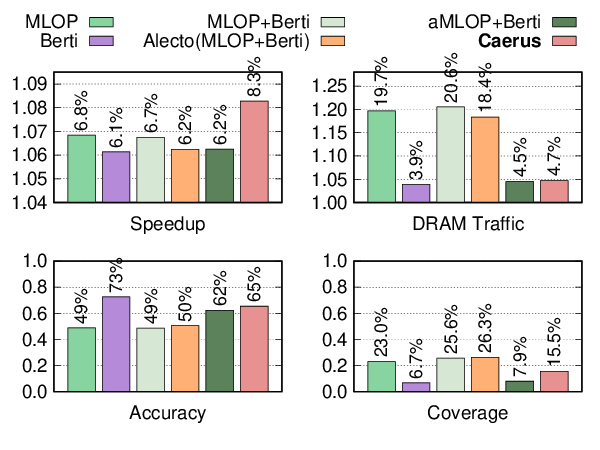}
    \caption{\revision{Caerus compared with combinations of global and PC-local offset L1D Prefetchers in SPECspeed 2017.} }
    \label{fig:variants}
\end{figure}

\revision{
\Cref{fig:variants} shows how Caerus compares to a composite of global and PC-local offset prefetching strategies. MLOP and Berti running simultaneously in the L1D does not perform as well as MLOP on its own. Although coverage is slightly higher, the extra DRAM traffic results in a slight net performance loss. Using the prefetcher selection algorithm Alecto~\cite{alecto}, applied to Berti and MLOP combined, also results in reduced performance gains. While Berti does not see much filtering from Alecto due to its high accuracy, MLOP sees extensive filtering but not always in a way that is helpful. While workloads like cactuBSSN are improved by Alecto removing inaccurate prefetches (from 14.2\% to 21.4\% speedup, better than MLOP's 13.9\% and Berti's 0.0\%, but still worse than Caerus' 30.9\%), performance meaningfully drops on many workloads including bwaves (15.1\% to 5.5\%) and roms (27.1\% to 21.2\%) due to Alecto's PC-based input filtering. In a cross-PC setting, which is where MLOP (and Caerus) achieve additional coverage over Berti, inaccurate prefetches at PC \textit{A} may be caused by observations originating from PC \textit{B}. If training is then disabled at PC \textit{A}, PC \textit{B} will still be trained on, potentially causing inaccurate prefetches at PC \textit{C} instead.}

\revision{Using Caerus' local accuracy filter (\cref{ssec:accuracytracking}) to filter prefetches from MLOP (\textit{aMLOP}), in combination with Berti, considerably reduces DRAM traffic by contrast, but performance is still significantly below Caerus. Under per-PC accuracy filtering, the aggressive overlap-prone offsets in MLOP lose much of their performance advantage compared to Caerus' overlap-resistant next-best offsets. Conversely, we demonstrate that filtering does not lower performance for Caerus (\cref{sec:ablation}).}

\revision{This demonstrates that Caerus is not merely a combination of Berti and MLOP's best parts, nor is it simply that combination with added filtering to MLOP or both. 
As previously described in \cref{sec:caerus}, Caerus bears far more resemblance to an evolved BOP than either previous state-of-the-art mechanisms. Though accuracy filtering is key to achieving Caerus' low DRAM traffic (\cref{sec:ablation}), this filtering requires our novel methods of overlap-resistant offset derivation in order to perform well.
}

\begin{figure}[t]
    \centering
    \begin{subfigure}[t]{\columnwidth}
        \centering
        \includegraphics[width=\linewidth]{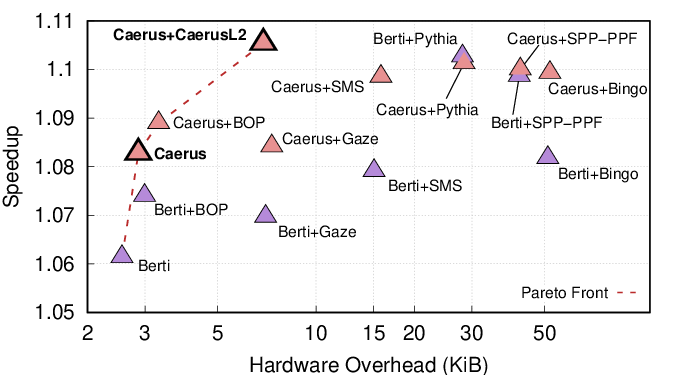}
        \label{fig:multilevel-storage}
    \end{subfigure}
    \begin{subfigure}[t]{\columnwidth}
        \centering
        \includegraphics[width=\linewidth]{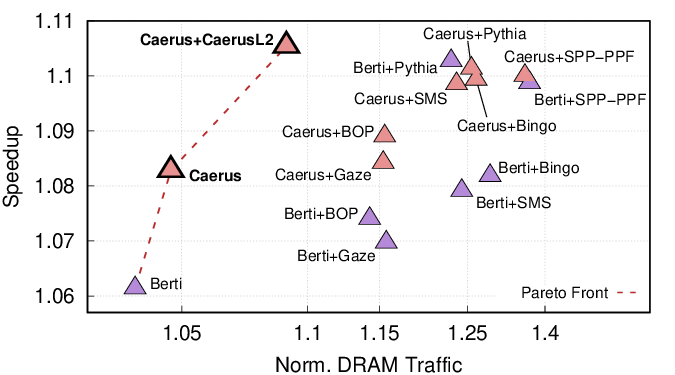}
        \label{fig:multilevel-dram-speedup}
    \end{subfigure}
        \caption{Speedup in SPECspeed 2017 vs. overhead for Caerus and Berti in the L1D combined with various state-of-the-art L2 prefetchers. Caerus+CaerusL2 achieves the highest performance out of any combination of prefetchers, and at comparatively low hardware and DRAM traffic overhead.}
    \label{fig:multi-level}
\end{figure}

\subsection{Multi-Level Prefetching}\label{ssec:l2}

\Cref{fig:multi-level} shows how both Berti and Caerus combine with commonly used L2 prefetchers: BOP~\cite{michaud_best-offset_2016}, Gaze~\cite{gaze}, SPP-PPF~\cite{SPP,PPF} and Pythia~\cite{Pythia}, all at default settings from their original papers. \revision{We also evaluate SMS~\cite{SMS-ISCA06} and Bingo~\cite{bingo}, sized at 12.5KiB and 48.6KiB respectively.} Caerus + CaerusL2 is the best performing combination while incurring considerably less hardware and DRAM traffic overhead compared to all close contenders. 

Berti gains considerable performance even from the addition of BOP, because it cannot handle cross-PC offsets alone. However, this combination does not outperform Caerus on its own and incurs significantly more DRAM traffic. Caerus gains only slightly from the addition of Gaze and BOP because it already effectively covers cross-PC patterns. \revision{SMS and Bingo bring moderately more performance gains by capturing non-linear spatial patterns, but  Caerus in the L1D alone still outperforms combinations of SMS and Bingo with Berti.} SPP-PPF and Pythia add significant performance to both Berti and Caerus. Nevertheless, Caerus combined with \textit{itself}, configured as an L2 prefetcher\footnote{To account for the L2C's greater capacity, CaerusL2's RR table is sized at 64 sets each with 16 entries (4.04 KiB overall prefetcher state). 
\revision{Deployed in the L2C without an L1D prefetcher, CaerusL2 achieves a 10.0\% speedup in SPECspeed 2017, again outperforming all competitors, for example SMS (6.84\%), Bingo (7.36\%), SPP-PPF (9.09\%), and Pythia (9.10\%). }}, outperforms any other combination. Compared with Berti+Pythia (the second-highest performing combination), Caerus+CaerusL2 brings better performance while incurring only a fraction of the DRAM traffic (8.9\% vs. 22.7\%) and requiring significantly less hardware cost (6.90KiB vs. 28.05KiB).

\subsection{\revision{Multi-Core Performance Analysis}}

\begin{figure}[t] 
    \centering
    \includegraphics[width=\columnwidth]{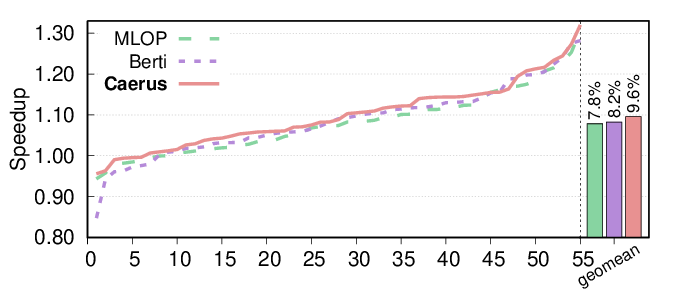}
    \caption{Speedup of Caerus, MLOP, and Berti across 55 randomly selected 8-core heterogeneous workload mixes from all benchmark suites (sorted by speedup). }
    \label{fig:multicore-hetero}
\end{figure}

\begin{figure}[t] 
    \centering
    \includegraphics[width=\columnwidth]{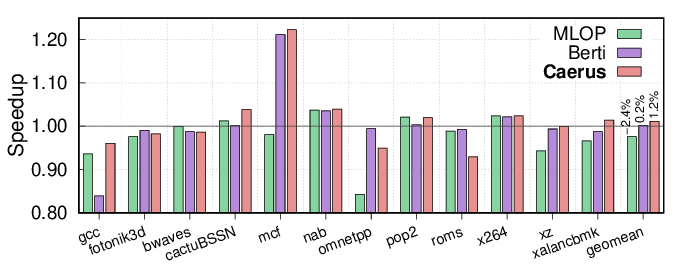}
    \caption{Speedup of Caerus, MLOP, and Berti in 8-core SPECspeed 2017 homogeneous workloads that show a performance delta of larger than 1\%. }
    \label{fig:multicore-homo}
\end{figure}

\revision{
\Cref{fig:multicore-hetero} and \Cref{fig:multicore-homo} show 8-core heterogeneous and homogeneous multi-core performance of Caerus, MLOP, and Berti. 
Caerus shows consistent speedups in heterogeneous workloads, achieving a 9.6\% geomean speedup compared to Berti at 8.2\% and MLOP at 7.8\%. MLOP's inefficient use of DRAM traffic is exacerbated in this multi-core setting causing it to be the lowest performer in 31 out of 55 mixes. In homogeneous SPECspeed 2017 workloads, while the benefits of all prefetchers are limited because these workloads are highly DRAM-bound, Caerus achieves a geomean performance improvement of 1.0\% over Berti and 3.6\% over MLOP. %
}

\subsection{Parameter Sensitivity}
\label{ssec:sensitivity}

\begin{figure}[t]
    \centering
    \includegraphics[width=\columnwidth]{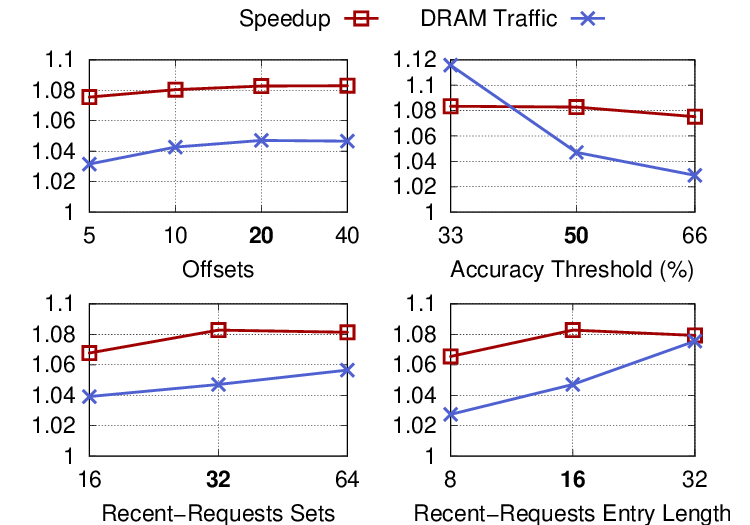}
    \caption{Sensitivity study of Caerus parameters on SPECspeed 2017 (default parameters indicated in \textbf{bold}).}
    \label{fig:sensitivity}
\end{figure}

\Cref{fig:sensitivity} shows Caerus' sensitivity to its various parameters. Performance is only weakly affected by increasing global offsets past 10, due to Caerus already avoiding overlap in the offsets trained and immediately retraining offsets when needed. Lowering or increasing the accuracy threshold (implemented via biasing the saturating counters~\cite{ainsworth_triangel_2024}) is able to trade off speedup and traffic, with a 66\% threshold lowering traffic to just 2.9\% while still outperforming Berti. Having fewer sets means more 4KiB regions end up in the same entry concurrently, reducing the lifetime of RR elements and thus lowering training coverage. Increasing the length of each shift register in the RR sets increases aggression by making the bar for accuracy lower, producing a sweet spot based on L1D capacity. When using 16 RR sets with 8 entries each (at 1.88 KiB total state versus Berti's 2.55KiB), Caerus still outperforms Berti while consuming 1.5\% less DRAM traffic. 

\begin{figure}[t]
    \centering
    \includegraphics[width=\columnwidth]{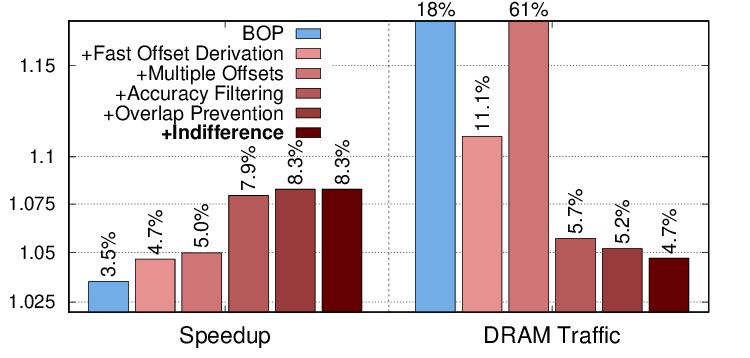}
    \caption{Ablation Study. Starting with BOP and progressively adding Caerus' features on SPECspeed 2017.}
    \label{fig:ablation}
\end{figure}

\subsection{Ablation Study} \label{sec:ablation}

\Cref{fig:ablation} shows the contribution of each of Caerus' core components. A version of Caerus that trains one offset and has no accuracy filtering still outperforms BOP. This is because the same-region \textit{fast offset derivation} training method (\cref{ssec:history}) greatly increases Caerus' training speed and thus its ability to adapt to access-pattern changes. Adding \textit{multiple offsets} slightly increases performance but comes with untenable DRAM traffic. This traffic is massively cut down by \textit{accuracy filtering} (\cref{ssec:accuracytracking}), allowing a multiple-offset approach to give substantial benefits. \textit{Preventing overlap} (\cref{ssec:overlap}) between offsets adds a final speedup while reducing DRAM traffic. The addition of \textit{indifference} (\cref{ssec:accuracytracking}) distinguishes prefetches that will be L2 hits instead of L1D hits, reducing DRAM traffic particularly in \bench{leela} and \bench{perlbench} by selectively reducing aggression but keeping consistent performance.

\section{Related Work}

Prefetching is a widely researched topic. Surveys of the full area have been performed by Mittal~\cite{1198_Mittal} and by Wenisch and Falsafi~\cite{1197_Falsafi}. We document the most relevant topics here.

\subsection{Offset Prefetchers}

Sandbox prefetching~\cite{Sandbox} introduced the concept of prefetching by offsets: address-deltas that would have been accurate based on looking at a wide selection of recent history rather than just the last access (as is typical in e.g., stride prefetchers~\cite{stride}). Best-offset prefetching~\cite{michaud_best-offset_2016,michaud_dpc2} formalizes many of the details and also introduces the concept of timely offsets: the best offset is a historical address-delta that is able to bring in the highest number of memory accesses where the data arrives back to the cache before they are demand-requested.
MLOP extends further to produce a set of offsets varying in timeliness~\cite{shakerinava_multi-lookahead_2019}. 

Berti generalizes timely offsets by training on a PC-local viewpoint to choose distinct sets of offsets per-PC, allowing much more precise prefetching than BOP, but missing out on cross-PC potential.  Berto~\cite{berto} uses a more efficient method to track timeliness, and increases the size of other structures to better fit server workloads. Chimera~\cite{he_chimera_2024} and Hyperion~\cite{Hyperion} introduce a set of Berti-style prefetchers, training on local groups other than PCs, including region-local best offsets. BertiGO~\cite{Singh2026BertiGO} and Global Berti (GBerti)~\cite{Posluns2026GlobalBerti} took first and second place at the 4th Data Prefetching Championship~\cite{dpc4_2026} respectively. The former adds filtering and multi-PC histories to Berti, and the latter adds a more limited version of global offsets to Berti than Caerus sees, by only looking at the last access at all PCs (i.e., a column access to the history table in addition to Berti's row access), where Caerus sees all recent accesses of all PCs within a page.

\subsection{Other Prefetch Techniques}

Stride prefetchers~\cite{stride}, which learn scenarios at a PC-local level where the difference between address pairs are constant, are widely deployed~\cite{bruce_arm_2023} but less general than offset prefetching. Spatial Memory Streaming~\cite{SMS-ISCA06,Dspatch,bingo}, where patterns that repeat across different regions (e.g., pages) are learned, are also widely deployed~\cite{bruce_arm_2023}. Temporal prefetchers~\cite{Triage-MICRO19,ainsworth_triangel_2024,MISC-ISCA19,ISB,GHB}, which store large amounts of past historical addresses rather than arithmetic relations between them, have recently become viable~\cite{bruce_arm_2023}. Addresses can be used to correlate patterns more generally, such as the Signature Path Prefetcher~\cite{SPP} which learns sequences along with confidence values in those sequences.  Neural-inspired techniques are also becoming gradually more implementable~\cite{Hierarchical,NewForm}. Other prefetchers learn patterns based on data itself rather than addresses: these include run-ahead mechanisms~\cite{runahead,DVR,SVR,1263_Naithani}, which precalculate segments of programs to learn their memory accesses, and the Indirect Memory Prefetcher~\cite{IMP}, which spots use of indexed values in arrays. 

\subsection{Prefetcher Control}

Caerus is not the first prefetching technique to throttle prefetches via accuracy-centric mechanisms. Feedback-directed prefetching~\cite{Feedback} measures prefetcher accuracy and adapts aggression, for example by increasing or decreasing the multiplicative degree of stride or next-line prefetchers (e.g., issuing for a stride S and base address A, $A+S,A+2S$ and $A+3S$). Most offset prefetchers, including BOP, Caerus and Berti, issue prefetches at multiplicative degree 1 even if Caerus and Berti can potentially issue multiple distinct offsets at once. Since they already handle timeliness through direct tracking rather than through aggression control, feedback-directed prefetching is harder to directly apply. Perceptron prefetch filtering~\cite{PPF} potentially allows richer filtering to be built on top of Caerus' accuracy table, and we leave this exploration for future work. Aggression control has also been explored at a coarser granularity to control groups of prefetchers, such as in IPCP~\cite{IPCP} and Alecto~\cite{alecto}. Caerus' decisions are more fine-grained, working within the prefetcher itself, adapting the core prefetch algorithm to the use of accuracy control, and it can measure accuracy without actually issuing prefetches by using sandbox techniques~\cite{Sandbox}, meaning it can drop aggression to 0 when needed. Properties other than accuracy can also be considered in whether to issue prefetches, such as criticality~\cite{CLIP}.

\section{Conclusion}

We have introduced Caerus, an offset prefetcher that thinks globally and acts locally. It achieves high performance with sufficiently low DRAM-traffic and hardware overhead for real-world use in systems of all scales. It outperforms the state-of-the-art offset prefetcher Berti with similar hardware size while avoiding Berti's expensive timestamp tags on L1D cache lines. 
We hope that its combination of practicality, performance, and suitability across many cache levels will enable the rapid deployment of Caerus.

\newpage

\section*{Artifact Appendix}

\subsection{Abstract}

This artifact contains the source code and simulation framework used to evaluate Caerus. The repository contains: 
\begin{itemize}
    \item Source code for Caerus. 
    \item Source code for the Stride prefetcher, BOP~\cite{michaud_best-offset_2016}, MLOP~\cite{shakerinava_multi-lookahead_2019}, Chimera~\cite{he_chimera_2024}, and Berti~\cite{navarro-torres_berti_2022}. 
    \item Modified ChampSim framework used to evaluate all presented prefetchers. 
    \item Scripts to download SPECspeed 2017 traces, run simulations, and collect results.
\end{itemize}

This artifact reproduces the data required to plot \Cref{fig:speedup17}, \Cref{fig:benches-dram}, and the Accuracy / Coverage portion of \Cref{fig:benches-acccov} for SPECspeed 2017. 

\subsection{Artifact check-list (meta-information)}

{\small
\begin{itemize}
  \item {\bf Data set: } SPECspeed 2017, download using provided scripts
  \item {\bf Metrics: } Speedup, DRAM Traffic, Coverage, Accuracy
  \item {\bf Output: } Table with per-benchmark metrics for each prefetcher
  \item {\bf Experiments: } Baseline (no prefetching), Caerus. Optional: Stride, BOP, MLOP, Chimera, Berti
  \item {\bf How much disk space required (approximately)?: } 30GB 
  \item {\bf How much time is needed to prepare workflow (approximately)?: } 50 minutes to download workload traces 
  \item {\bf How much time is needed to complete experiments (approximately)?: } 2 hours for reproducing main results (baseline + Caerus). [Optional] 7 hours for reproducing results for baseline + all presented prefetchers. Tested on a Xeon Platinum 8168 machine using 40 cores. 
  \item {\bf Publicly available?: } Yes 
  \item {\bf Code licenses (if publicly available)?: } Apache 2.0
  \item {\bf Archived (provide DOI)?: } \url{https://doi.org/10.5281/zenodo.21512560}
\end{itemize}
}

\subsection{Description}

\subsubsection{How to access}

The source code is available on GitHub (\url{https://github.com/jw38176/Caerus_Artifact/}) and Zenodo (\url{https://doi.org/10.5281/zenodo.21512560}). 

\subsubsection{Software dependencies}

Installing and running this artifact requires the following software dependencies:    
    \texttt{Python 3.9+}, 
    \texttt{g++} with C++17 support, 
    \texttt{make} 
    and \texttt{git}.

The artifact is tested with the following system specs:
    Fedora 44, 
    Linux Kernel 7.0.13, 
    GCC 16.1.1 and
    Python 3.14.6.

\subsubsection{Data sets}

This artifact contains a script to download SPECspeed 2017 traces used in DPC3~\cite{champsim_traces_speccpu_2025}, hosted on \url{https://dpc3.compas.cs.stonybrook.edu/champsim-traces/speccpu/}. 

This artifact can also be easily modified to reproduce all other evaluated benchmarks. The SPEC CPU2006 traces are hosted on the same archive as the SPECspeed 2017 traces. The Ligra traces can be found at \url{https://zenodo.org/records/14267978} and PARSEC traces at \url{https://zenodo.org/records/14268119}. 

\subsection{Installation}

Use the following steps to download the source code, install dependencies for ChampSim, and download the required traces: 

\newpage

\begin{minted}[fontsize=\footnotesize]{bash}
# Clone the repository
git clone \
    https://github.com/jw38176/Caerus_Artifact.git
cd Caerus_Artifact

# Build champsim dependencies
git submodule update --init
vcpkg/bootstrap-vcpkg.sh
vcpkg/vcpkg install

# Download SPECspeed 2017 traces 
python3 scripts/download_spec2017_traces.py
\end{minted}

\subsection{Experiment workflow}

In the commands below, change the parameter $N$ to run $N$ experiments in parallel.

\subsubsection{To reproduce Caerus}

\begin{minted}[fontsize=\footnotesize]{bash}
# Run baseline (no prefetching) and Caerus
python3 scripts/run_experiments.py no -j N
python3 scripts/run_experiments.py caerus -j N
\end{minted}

\subsubsection{[OPTIONAL] To reproduce all prefetchers}

\begin{minted}[fontsize=\footnotesize]{bash}
# [OPTIONAL] Run baseline and all prefetchers
for pref in no ip_stride bop mlop \
            chimera berti caerus; do
    python3 scripts/run_experiments.py "$pref" -j N
done
\end{minted}

\subsection{Evaluation and expected results}

The \mintinline{bash}{scripts/print_results.py} script produces a table of the per-benchmark and geomean results for Speedup, DRAM Traffic, Accuracy, and Coverage. Using geomean Speedup as an example:  

\subsubsection{Reproducing Caerus}

\begin{minted}[fontsize=\footnotesize]{bash}
# Print Caerus results 
python3 scripts/print_results.py caerus

# Results should look like: 
SPEEDUP:
  caerus      1.0828
\end{minted}

\subsubsection{[OPTIONAL] Reproducing all prefetchers}

\begin{minted}[fontsize=\footnotesize]{bash}
# [OPTIONAL] Print all prefetchers results 
python3 scripts/print_results.py ip_stride bop \
  mlop chimera berti caerus

# Results should look like: 
SPEEDUP:
  ip_stride   1.0319
  bop         1.0354
  mlop        1.0684
  chimera     1.0340
  berti       1.0614
  caerus      1.0828
\end{minted}

\subsection{Experiment customization}

Caerus, along with other prefetchers, are implemented in the directory \mintinline{bash}{prefetcher/caerus}. You can change the parameters of Caerus in \mintinline{bash}{caerus_parameters.h}. You can also change the parameters of the simulated processor and cache in \mintinline{bash}{params/baseline.json}. 

\subsection{Methodology}

Submission, reviewing and badging methodology:

\begin{itemize}
  \item \url{https://www.acm.org/publications/policies/artifact-review-and-badging-current}
  \item \url{https://cTuning.org/ae}
\end{itemize}

\bibliographystyle{IEEEtranS}
\bibliography{refs, refs2}

\end{document}